\documentclass[english]{article}
\usepackage[LGR,T1]{fontenc}
\usepackage[latin9]{inputenc}
\usepackage{geometry}
\usepackage{textcomp}
\usepackage{amsmath}
\usepackage{graphicx}
\usepackage{setspace}
\usepackage[authoryear]{natbib}
\makeatletter

\DeclareRobustCommand{\greektext}{%
  \fontencoding{LGR}\selectfont\def\encodingdefault{LGR}}
\DeclareRobustCommand{\textgreek}[1]{\leavevmode{\greektext #1}}
\ProvideTextCommand{\~}{LGR}[1]{\char126#1}

\newcommand{\lyxdot}{.}

\newcommand{\lyxaddress}[1]{
\par {\raggedright #1
\vspace{1.4em}
\noindent\par}
}

\makeatother

\usepackage{babel}
\begin{document}

\title{On the astronomical origin of the Hallstatt oscillation found in
radiocarbon and climate records throughout the Holocene }

\author{Nicola Scafetta$^{1\dagger}$, Franco Milani$^{2}$, Antonio Bianchini$^{3,4}$,
Sergio Ortolani$^{3,4}$}
\maketitle

\lyxaddress{$^{1}$Department of Earth, Environmental and Resources Science,
Università degli Studi di Napoli Federico II, Napoli, Italy}

\lyxaddress{$^{2}$Astronomical Association Euganea, Padova, Italy}

\lyxaddress{$^{3}$Department of Physics and Astronomy, University of Padova,
Italy}

\lyxaddress{$^{4}$INAF-Osservatorio Astronomico di Padova, Vicolo dell'Osservatorio
5, I-35122 Padova, Italy}

\lyxaddress{$^{\dagger}$email: nicola.scafetta@unina.it}
\begin{abstract}
An oscillation with a period of about 2100-2500 years, the Hallstatt
cycle, is found in cosmogenic radioisotopes ($^{14}C$ and $^{10}Be$)
and in paleoclimate records throughout the Holocene. This oscillation
is typically associated with solar variations, but its primary physical
origin remains uncertain. Herein we show strong evidences for an astronomical
origin of this cycle. Namely, this oscillation is coherent to a repeating
pattern in the periodic revolution of the planets around the Sun:
the major stable resonance involving the four Jovian planets - Jupiter,
Saturn, Uranus and Neptune - which has a period of about $p=2318$
yr. Inspired by the Milankovi{\'c}'s theory of an astronomical origin
of the glacial cycles, we test whether the Hallstatt cycle could derive
from the rhythmic variation of the circularity of the solar system
disk assuming that this dynamics could eventually modulate the solar
wind and, consequently, the incoming cosmic ray flux and/or the interplanetary/cosmic
dust concentration around the Earth-Moon system. The orbit of the
planetary mass center (PMC) relative to the Sun is used as a proxy.
We analyzed how the instantaneous eccentricity vector of this virtual
orbit varies from 13,000 B. C. to 17,000 A. D.. We found that it undergoes
kind of pulsations as it clearly presents rhythmic contraction and
expansion patterns with a 2318 yr period together with a number of
already known faster oscillations associated to the planetary orbital
stable resonances. There exists a quasi $\pi/2$ phase shift between
the 2100-2500 yr oscillation found in the $^{14}C$ record and that
of the calculated eccentricity function. Namely, at the Hallstatt-cycle
time scale, a larger production of radionucleotide particles occurs
while the Sun-PMC orbit evolves from more elliptical shapes ($e\approx0.598$)
to more circular ones ($e\approx0.590$), that is while the orbital
system is slowly imploding or bursting inward; a smaller production
of radionucleotide particles occurs while the Sun-PMC orbit evolves
from more circular shapes ($e\approx0.590$) to a more elliptical
ones ($e\approx0.598$), that is while the orbital system is slowly
exploding or bursting outward. Since at this timescale the PMC eccentricity
variation is relatively small ($e=0.594\pm0.004$), the physical origin
of the astronomical 2318 yr cycle is better identified and distinguished
from faster orbital oscillations by the times it takes the PMC to
make pericycles and apocycles around the Sun and the times it takes
to move from minimum to maximum distance from the Sun within those
arcs. These particular proxies reveal a macroscopic 2318 yr period
oscillation, together with other three stable orbital resonances among
the outer planets with periods of 159, 171 and 185 yr. This 2318 yr
oscillation is found to be spectrally coherent with the $\varDelta^{14}C$
Holocene record with a statistical confidence above 95\%, as determined
by spectral analysis and cross wavelet and wavelet coherence analysis.
At the Hallstatt time scale, maxima of the radionucleotide production
occurred when, within each pericycle-apocycle orbital arc, the time
required by the PMC to move from the minimum to the maximum distance
from the Sun varies from about 8 to 16 years while the time required
by the same to move from the maximum to the minimum distance from
the Sun varies from about 7 to 14 years, and vice versa. Thus, we
found that a fast expansion of the Sun-PMC orbit followed by a slow
contraction appears to prevent cosmic rays to enter within the system
inner region while a slow expansion followed by a fast contraction
favors it. Similarly, the same dynamics could modulate the amount
of interplanetary/cosmic dust falling on Earth. Indeed, many other
stable orbital resonance frequencies (e.g. at periods of 20 yr, 45
yr, 60 yr, 85 yr, 159-171-185 yr, etc.) are found in radionucleotide,
solar, aurora and climate records, as determined in the scientific
literature. Thus, the result supports a planetary theory of solar
and/or climate variation that has recently received a renewed attention.
In our particular case, the rhythmic contraction and expansion of
the solar system driven by a major resonance involving the movements
of the four Jovian planets appear to work as a gravitational/electromagnetic
pump that increases and decreases the cosmic ray and dust densities
inside the inner region of the solar system, which then modulate both
the radionucleotide production and climate change by means of a cloud/albedo
modulation. 
\\
\textbf{Cite}: Scafetta, N., Milani, F., Bianchini, A., Ortolani,
S.: On the astronomical origin of the Hallstatt oscillation found
in radiocarbon and climate records throughout the Holocene. Earth-Science
Reviews 162, 24\textendash 43, 2016. http://dx.doi.org/10.1016/j.earscirev.2016.09.004
\end{abstract}

\section{Introduction}

Cosmic rays continuously collide with the Earth's atmospheric molecules
fragmenting their nuclei and producing neutrons. The collisions of
thermal neutrons with nitrogen nuclei ($_{7}^{14}N$ made of 7 protons
and 7 neutrons) give origin to cosmogenic radioisotopes ($_{6}^{14}C$
made of 6 protons and 8 neutrons) according to the following reaction:
$n+{}_{7}^{14}N{}\rightarrow{}_{6}^{14}C+p$. $^{14}C$ rapidly reacts
with oxygen to produce $CO_{2}$ and, as such, is absorbed by biological
organisms such as trees and marine corals. Since their formation and
their capture by biological systems, $^{14}C$ atoms undergo a radioactive
beta decay into stable $^{14}N$ atoms with a half-life time of 5730
years according to the following reaction: $ _{6}^{14}C\rightarrow{}_{7}^{14}N+e^{-}+\bar{\nu}_{e}$.
Thus, by measuring the percent of $^{14}C$ atoms present in a specific
organic material, if the age of the latter can be timed independently,
it is possible to determine the $^{14}C$ atmospheric concentration
of the past. Determining this time series is astronomically and geophysically
important because $^{14}C$ concentration variations are a direct
consequence of changes in the intensity of the cosmic ray flux reaching
the Earth, in solar magnetic activity \citep{Bard1997,Stuiver1980},
in the Earth\textquoteright s dipole moment \citep{Elsasser,Lal,OBrien}
and in a number of parameters regulating the radiocarbon exchange
system \citep{Oeschger,Siegenthaler,Stocker,Goslar}. 

Several experimental evidences demonstrate that $^{14}C$ concentration
varies in time \citep[e.g. ][and many others]{Damon1986,Kromer}.
\citet{Bray}, using short records, and later \citet{Houtermans}
using records spanning throughout the Holocene (since 10,000 B. C.),
noted that $^{14}C$ concentration has changed cyclically with the
longest certain period being about 2100-2500 years long. Longer oscillations
could be present, but the record was too short to detect them. This
period is known in the literature as the Hallstatt cycle \citep{Vasiliev}
because its cooling minimum occurred before the Maunder Minimum {[}1645:1715{]}
happened about 2800 years ago during a late Bronze - early Iron cultural
transition in an Austrian archaeological site located at Hallstatt.
Other major oscillations found in the $^{14}C$ concentration records
have periods of about 900-1050 year \citep{Bond,Kerr,Scafetta2012a,Davis}
known in the literature as the Eddy cycle \citep[cf.][]{McCracken2013}
and a 208-year cycle known in the literature as the Suess or de Vries
cycle \citep{Sonett1984}. The presence of fundamental harmonics in
radiocarbon records have been confirmed by numerous studies \citep{Abreu,Damon1986,Damon1988,Damon1990,Damon1992,Damon1992b,McCracken2013,Vasiliev1998,Vasiliev}. 

An oscillation with a period of about 2100-2500 years has been found
also in a number of paleoclimate records and/or events throughout
the 12,000 years of the Holocene \citep[e.g.: ][]{Levina1993,OBrien1995}.
For example, it was found in the $\delta^{18}O$ concentration measured
in ice cores and in deep-see cores with high sedimentation rates \citep{Pestiaux}.
Dendroclimatological considerations have also demonstrated that the
Little Ice Age (1500\textendash 1800 year A.D.), the Hallstattzeit
cold epoch (750\textendash 400 year B.C.) and the earlier cold epoch
(3200\textendash 2800 year B.C.) are separated by 2100\textendash 2500
years \citep[p. 378]{Damon1992}. Given the evident correlation and
synchronicity between the 2100-2500 year oscillation found in the
$^{14}C$ concentration record and in a number of paleoclimatic data,
all these records must be linked together. 

Climate variations and ocean/air ventilation changes could also modulate
the production of $^{14}C$ concentration. However, this interpretation
leaves unanswered the question of why the climate would oscillate
with a 2100-2500 year cycle. $^{14}C$ concentration could also vary
because of changes in the Earth\textquoteright s magnetic field shielding
the Earth from incoming cosmic rays \citep[e.g. ][]{Damon1986}. However,
changes of the Earth\textquoteright s dipole field may be too weak
to cause the 2100-2500 year oscillation in the radiocarbon records
\citep[cf.: ][]{Creer,Damon1992}.

Also cosmic rays, which directly form cosmogenic radioisotopes, influence
the Earth's climate. Indeed, numerous empirical evidences and theoretical
considerations have pointed out that cosmic rays can contribute to
the formation of clouds and, therefore, modulate the Earth's albedo
by ionizing the atmosphere \citep[e.g.: ][]{Kirby,Svensmark,Svensmark2,Svensmark2012,Tinsley},
although cosmic rays alone may not explain the full amount of atmospheric
precipitation variation. The existence of an astronomical origin of
the involved mechanisms would also be supported by the finding that
variations in $^{14}C$ concentration are correlated with the solar
system\textquoteright s galactic motion and imprinted in the Phanerozoic
climate over the last 600 million years \citep[e.g.: ][]{Shaviv}.

Several authors have concluded that the observed 2100-2500 year oscillation
both in the cosmogenic radioisotope records and in the climate records
has a solar origin \citep[e.g.: ][]{Dergachev,Hood,Hoyt}. Indeed,
$^{14}C$ records, as well as $^{10}Be$ records reproduce features
present in the sunspot number records such as the Maunder and Dalton
solar minima, and other solar directly observed patterns \citep[cf.: ][]{Adolphi,Bard1997,Bard,Scafetta2012a,Steinhilber,Usoskin}.
However, these considerations still do not explain why solar activity
should vary with a 2100-2500 year oscillation. Indeed, this oscillation
might also be forced on the system.

In any case, even if cosmic rays are one of the drivers of climate
change, one should explain why they are modulated by a 2100-2500 years
periodicity. The origin of this can be of three kinds: astronomical,
solar, or Earth\textquoteright s endogenous. \citet{Gregori2002}
suggested that the encounters of the Solar System with clouds of interstellar
matter modulate solar physics, hence its activity, and also its release
of solar wind. The Earth, with its magnetosphere, captures a fraction
estimated at $\sim0.5\times10^{-9}$ of the surface, at 1 AU, of the
expanding solar corona. That is, this is a very tiny fraction of the
whole volume of the out-flowing solar wind. Similarly, the solar system
is presumably capturing a very tiny fraction of the clouds of interstellar
matter. These records are expected to be erratic and/or, on multi-million
year time scale, they can also be modulated by the movement of the
solar system within the galaxy \citep[cf.: ][]{Gregori2002,Shaviv}.
However, at the shorter time scales the incoming dust flux might be
also modulated by the internal oscillating dynamics of the solar system.
Hence, a long term solar modulation could be only indicative of a
galactic modulation of solar physics.

Herein we hypothesize that the 2100-2500 yr oscillation in the radiocarbon
records has an astronomical origin, and search whether an astronomical
record clearly manifests such an oscillation. In this regard, \citet{Charvatova}
was the first in suggesting that the observed 2100-2500 yr oscillation
could be caused by the solar inertial motion, that is by the wobbling
of the Sun around the barycenter of the solar system due to the orbital
movements of its planets. She proposed a simplified model where the
2100-2500 yr oscillation had to be on average 2402.2 years long, which
would corresponds to the Jupiter/Heliocenter/barycenter alignments
(9.8855 x 243 = 2402.2 year). About the secular solar oscillations
\citet{Charvatova} showed that the inertial motion of the Sun varies
from a trefoil ordered state, where the orbital patterns nearly repeat
while rotating relative to the fixed stars, to a disordered one, where
the orbits show confused and chaotic patterns. The ordered cases correspond
to stable patterns correlated with historical solar maxima while the
disordered ones correlate with historical solar minima such as the
Wolf minimum (1280 to 1340), the Spörer minimum (1420 to 1570), the
Maunder minimum (1645 to 1715) and the Dalton minimum (1790 to 1820).
Based on these patterns, the current period (1985 to 2040) could yield
to a Dalton like minimum \citep[cf.: ][]{M=0000F6rner2015,Scafetta2010,Scafetta2012a}.
Moreover, again using a simplified model, \citet{Scafetta2012c} showed
that the conjunctions of Jupiter and Saturn, using their tropical
orbital periods, fully precess over a quasi 2400 year period. 

However, the above models were oversimplified as they neglected the
presence of the other planets: they could be unconvincing because
the 2100-2500 yr oscillation was merely implicit in calculations whose
physical meaning was hypothetical. In general, it could be argued
that it is physically unlikely that the solar inertial motion could
be the direct cause of a variation in the solar activity because the
Sun is in free-fall in it and should not feel it. A more realistic
hypothesis requires that the solar inertial motion is mathematically
linked to some physical mechanism yielding a variability in solar
magnetic activity and/or in the incoming cosmic ray flux. The solar
inertial motion could be just a proxy collecting the relevant information
about the dynamics of the solar system. In principle, the planetary
motion can produce gravitational and/or electromagnetic forcings directly
onto the Sun, interacting with its magnetic activity, and/or within
the heliosphere. In this way, it could be modulating the incoming
flux of cosmic rays as well as the concentration of the interplanetary/cosmic
dust around the Earth-Moon system \citep{Ermakov,Ollila}. Such forcing
should then maintain the imprinting of its origins and be synchronized
with some planetary resonances. 

Some physical mechanisms explaining a planetary modulation of solar
and climate activity are currently investigated \citep{Abreu,Scafetta(2012b),ScafettaandWillson(2013b),Wolff}.
A planetary origin of solar and climate oscillations, which has been
proposed since antiquity, does have numerous empirical evidences and
some preliminary explanations \citep[e.g.: ][]{Abreu,Charv=0000E1tov=0000E1,Cionco,Hung,Fairbridge,Fairbridge2,Jakubcov=0000E1,Jose,McCracken2013,McCracken2014,M=0000F6rner2013,M=0000F6rner2015,Mortari,Puetz,Salvador,Scafetta2010,Scafetta2012a,Scafetta(2012b),Scafetta2013,Scafetta2014,Scafetta2016,ScafettaWillson2013a,ScafettaandWillson(2013b),Sharp,Solheim,TanCheng,Tattersall,Tattersalb,Wilson}. 

The present work aims to provide a robust evidence that the 2100-2500
yr Hallstatt oscillation found both in cosmogenic radioisotopes and
in climate records throughout the Holocene has an astronomical origin
linked to a major recurrence of particular displacements of the planets
around the Sun. In the choice of the appropriate astronomical proxy,
we were inspired by the \citet{Milankovitch-2} theory that links
the $\sim100,000$ yr variation of the Earth's orbit eccentricity
to the glacial cycles of the past 1 million years. Thus, we hypothesize
that the Hallstatt cycle could derive from an expansion-contraction
rhythmic dynamics of the solar system driven by the rotation of the
planets that yields to a specific set of stable orbital resonances. 

The dynamics of the solar system circularity is well described by
the wobbling of the planetary mass center (PMC) orbiting the Sun,
which scales the wobbling of the Sun relative to the barycenter of
the solar system. We used the ephemeris of the Sun relative to the
barycenter to derive such a complex orbit. Then we used a proposed
Keplerian constant of motion, the eccentricity vector \citep[e.g.:][]{Mungan},
to evaluate how the instantaneous eccentricity of the orbit of the
PMC varies in time. Using this observable we demonstrate that the
solar system circularity pulses with a 2100-2500 period together with
a number of already known oscillations associated to the orbital periods
of the planets.

The situation of having several stable orbital resonances and orbital
proxies made of many harmonics is not surprising because of the complexity
of the solar system. It is, however, highly confusing for identifying
the possible physical origin of a specific oscillation. The evident
analogous is the theory of ocean tides where generic tidal generation
potentials deduced from the Sun's and Moon's orbits relative to the
Earth produce a very large number of tidal constituent waves \citep{Doodson,Melchior}.
These oscillations are differentiated in the literature with a very
long list of Darwin's symbols indicating their physical origin such
as the $N$ (lunar Saros) tidal wave, $Sa$ (solar annual) tidal wave,
$Mm$ (lunar monthly) tidal wave, $M2$ (principal lunar semidiurnal)
tidal wave, the $S2$ (principal solar semidiurnal) tidal wave, $N2$
(larger lunar elliptical semidiurnal) tidal wave, etc \citep{Darwin}.
Alternative and specific astronomical proxies are needed to well highlight
each of these oscillations because they have a different physical
origin and many of them are just small perturbations of the dominant
$M2$ and $S2$ waves. 

To better identify the physical origin of the astronomical 2100-2500
yr cycle, and to separate it from the fast and larger oscillations
associated to the orbits of the planets, we searched for more appropriate
astronomical proxies. We collected the times it takes the Sun to make
pericycles and apocycles and the times it takes the Sun to move from
its minimum to maximum distances from the barycenter within these
arcs. We show that these particular astronomical proxies reveal a
macroscopic 2100-2500 yr period oscillation perfectly coherent to
the Hallstatt oscillation found in the radionucleotide records with
a statistical confidence above 95\%. 

Finally, we briefly hypothesize the physical mechanisms involved in
the process suggesting that the pulses of the solar system could be
modulating the solar wind and by that the incoming cosmic ray flux
and the cosmic dust concentration surrounding the Earth \citep[cf: ][]{morner1996}. 

\begin{figure}[!t]
\centering{}\includegraphics[width=1\textwidth]{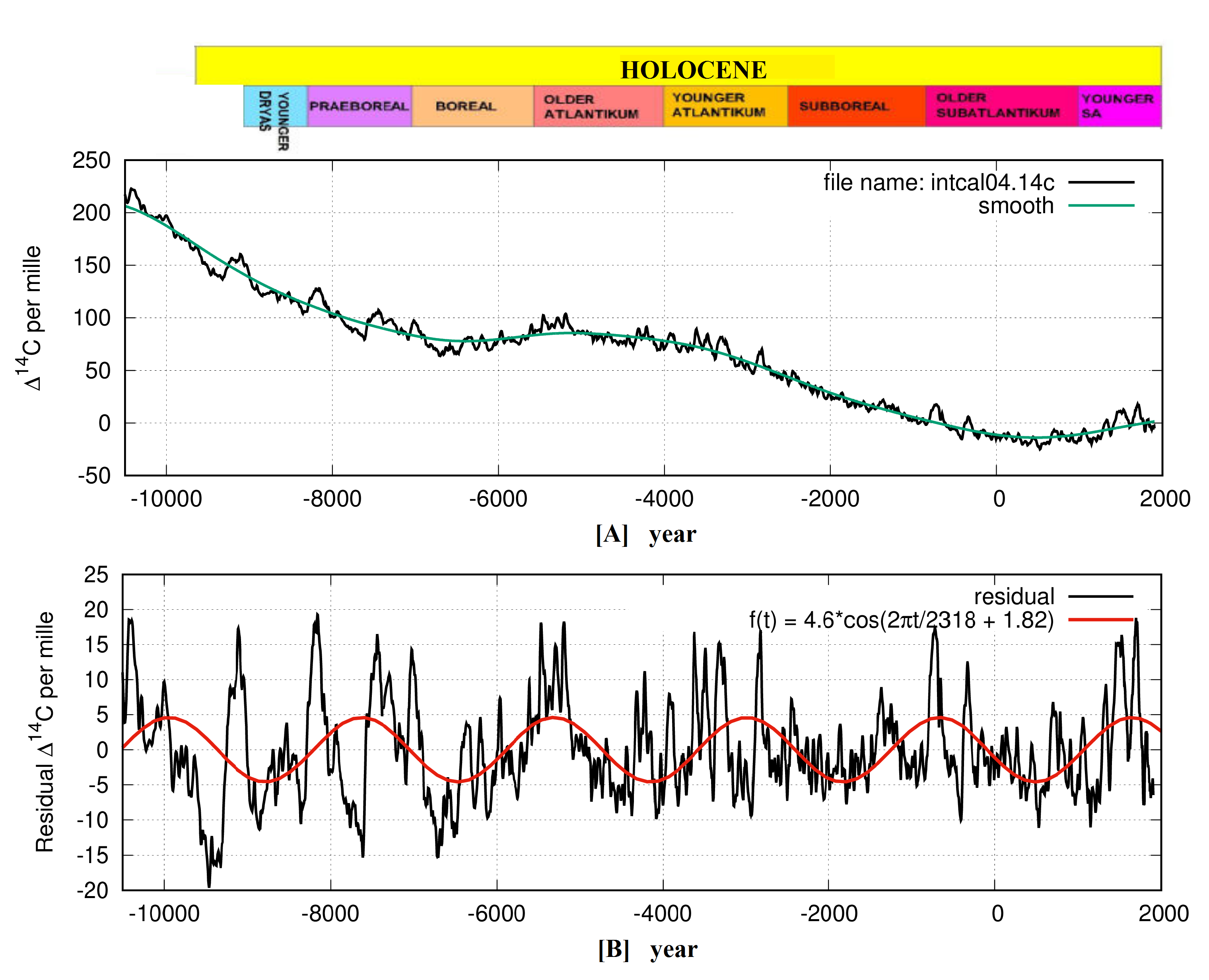}\caption{{[}A{]} $\varDelta^{14}C$ (\textperthousand ) record (black) throughout
the Holocene from -10500 B. C. to 1900 A. D. and its multi-millennial
smooth curve (green). {[}B{]} A residual signal obtained by detrending
the smooth curve from the data. The latter is a fit with a sinusoidal
function (red). This signal covers the period from 10500 B. C. to
1900 A. D. The figures report the name of the original file downloaded
from https://www.radiocarbon.org/IntCal04.htm}
\end{figure}

\section{The 2100-2500 yr Hallstatt cycle in the $\varDelta^{14}C$ nucleotide
record }

Figure 1A shows the $\varDelta^{14}C$ (\textperthousand ) record
(IntCal04.14c) from 10,500 B. C. to 1845 A. D. that covers the entire
Holocene \citep[data are from][]{Reimer}. This record was obtained
using dendrochronological dating and cross-checking from several millennia-long
tree-ring chronologies. The IntCal04.14c record extends for 26,000
years. The last 12,500 years are reported with a 5-year resolution
and the data derive mostly from tree-ring chronologies. For older
dates this record is made using mostly marine (e.g. coral) records
that have a lower resolution. The second component of the record,
the one obtained mostly from marine chronologies, is not used here
because it shows altered and very smooth patterns compared to the
tree-ring chronologies. This difference is very likely due to the
diverse physical properties of the two sets as one uses trees that
absorb carbon directly from the air where $^{14}C$ is produced by
cosmic rays, while the other uses corals that absorb carbon dissolved
in the water where it can remain diluted for very long times. 

Figure 1A also shows in green a gnuplot acsplines smooth curve that
captures the multi-millennial modulation of the record that appears
to be characterized by a trend plus an approximate 6,000-7,000 year
modulation, which is observed throughout the entire 26,000 years of
the original record. The physical origin of the trend and this long
oscillation are not discussed herein. Figure 1B shows the residual
signal obtained by detrending the smooth curve from the original data
from 10,500 B. C. to 1900 A. D.. Note that since 1900 the radionucleotide
record has been likely contaminated from increasing anthropogenic
$CO_{2}$ emissions and since 1950 because of atmospheric nuclear
bomb tests. 

The residual record depicted in Figure 1B is relatively similar to
the data made available in file Resid04\_2000.14c \citep{Reimer}
where a smoothing spline approximating a 2000-year moving average
was used \citep[cf: ][]{Stuiver}. However, our adopted filtering
is significantly smoother and better preserves the patterns at scales
shorter than 5000 years. This operation avoids artifacts that might
interfere with the Hallstatt cycle.

The residual record presents two major oscillations at about 900-1050
yr period (the Eddy cycle) and at 2100-2500 year period (the Hallstatt
cycle): see the periodogram in Figure 10. The quasi millennial oscillation
found in nucleotide record has been extensively studied in the literature.
It was found coherent with a quasi-millennial climate large oscillation
\citep[e.g.: ][]{Bond,Kerr} and was reconstructed with a combination
of Jupiter-Saturn tidal induced oscillations and the 11-year sun-spot
oscillation \citep[e.g.: ][]{Scafetta(2012b),Scafetta2014}. Herein,
we will not focus on this oscillation either, but only on the longer
Hallstatt cycle.

We attempted to fit the residual record depicted in Figure 1B with
a harmonic curve of the type 
\begin{equation}
f(t)=A\cos(2\pi t/T+\phi)+C.
\end{equation}
 However, the statistical fit gives slightly different values according
to the chosen time interval. For example: from 6000 B. C. to 1900
A. D. it gives $T=2357\pm24$ yr; from 7,000 B. C. to 1,900 A. D.
it gives $T=2326\pm18$ yr; from 8,000 B. C. to 1,900 A. D. it gives
$T=2249\pm16$ yr; from 9,000 B. C. to 1,900 A. D. it gives $T=2311\pm15$
yr; and from 10,500 B. C. to 1,900 A. D. it gives $T=2402\pm14$ yr.
The harmonic function fit also gives a slightly variable phase according
to the fit interval of about $\phi=2\pm0.2$. 

Such a variable result is not surprising in analyzing experimental
geophysical, radionucleotide or paleoclimatic records spanning several
thousands years since these records are characterized by some uncertainty
both in the amplitude and in the timing of the data. Moreover, the
records may be influenced by different physical sources that could
induce a certain pattern variability. In any case, since the statistical
error of the periodogram associated to a spectral peak period is $\nabla p=\pm p^{2}/2L$,
where $L=12,500$ yr is the length of the record shown in Figure 1B,
an observed spectral peak period at the 2100-2500 yr time scale is
characterized by a spectral error of about is $\pm200$ yr, which
well covers the typical oscillation range found in radionucleotide
($^{14}C$ and $^{10}Be$) records regarding the Hallstatt cycle,
as reported in the scientific literature.

Given the above uncertainty, in Figure 1B we fit the record with an
harmonic function with period $T=2318$ yr for the reasons explained
in Section 3 and also because such period is nearly recovered by the
fit value from 9,000 B. C. to 1,900 A. D. that covers the Holocene
after the end of the last glacial period. 

The harmonic function depicted in Figure 1B uses the fit parameters
obtained during the period between 750 B. C. and 1750 A. D., which
covers the first Hallstatt oscillation observed in the radionucleotide
record. During this period the data are likely the most accurate of
the record because the uncertainty increases with time. The phase
shift of the harmonic function is $\phi=1.82$. Thus, the maximum
of the radiocarbon Hallstatt cycle corresponds to about 1645 A. D.,
namely the beginning of the Maunder solar Minimum {[}1645-1715{]},
which was the most significant solar minimum of the last millennium. 

As shown in Figure 1B, the depicted harmonic function well predicts
the previous Hallstatt maxima in the radionucleotide record that corresponded
to the coldest epochs of the Holocene occurred around these periods:
10,000 B. C. (Younger Dryas cooling onset), 7,500 B. C. (Early Holocene
cooling event), 5,300 B. C. (Boreal/Atlantic transition and precipitation
change), 3,000 B. C. (Mid-Holocene Transition.), 700 B. C. (Sub-Atlantic
Minimum that also yielded the Greek Dark Ages). Indeed, \citet{OBrien1995}
found that also the polar atmospheric circulation changes are regulated
by a Hallstatt oscillation throughout the Holocene. A cooling-warming
cycle of about 2100-2500 year is, indeed, observed throughout the
Holocene in numerous climate records \citep[e.g.: ][]{Bond,Marcott,Mayewski}.
Also a quasi millennial cycle is observed in climate and radionucleotide
records \citep[cf.:][]{Bond,Kerr,Scafetta2013,Scafetta2014}.

In any case, note that the adoption of a fit phase of about $\phi=2$
would induce a temporal shift of about 65 years relative to the chosen
harmonic depicted in Figure 1B, and the maximum of the 2318 year cycle
would fall in 1580 A. D., which is between the Sp{\"o}rer solar Minimum
{[}1450-1550{]} and the Maunder Minimum {[}1645-1715{]}. Thus, in
any case, the strong solar minimum of the 16th-17th centuries was
likely driven by the Hallstatt cycle. For the purpose of this paper
the difference between the two phases is nearly negligible. 

\section{The Jupiter-Saturn-Uranus-Neptune 2318-year stable resonance}

An important concept in celestial mechanics is that of orbital resonance.
A resonance occurs when two or more orbiting bodies can exert a regular,
periodic gravitational influence on each other. This happens when
their orbital periods are related by a ratio of small integers. Orbital
resonances greatly enhance the mutual gravitational influence of the
bodies and, therefore, of the space symmetry of an orbiting system
such as the heliosphere of the solar system. Well known examples of
orbital resonances are the 1:2:4 resonance of Jupiter's moons Ganymede,
Europa and Io, the 2:3 resonance between Pluto and Neptune, the various
resonances that regulate the asteroid belt etc.. Indeed, the entire
solar system appears to be synchronized by specific orbital resonances
\citep[cf. ][]{Scafettab,Tattersall}. Thus, we hypothesize that the
Hallstatt oscillation found in radionucleotide and climatic records
could be the result of a specific orbital resonance within the solar
system. All planets could be involved in the process but, because
of the length of the Hallstatt oscillation, it is more reasonable
to search a resonance that links the four Jovian planets: Jupiter,
Saturn, Uranus and Neptune.

A system of periods $T_{i}$ is said to be in linear resonant state
if there exists a set of small integer numbers $a_{i}$ such that: 

\begin{equation}
\frac{1}{T}=\left|\sum\frac{a_{i}}{T_{i}}\right|<\gamma,\label{eq:2}
\end{equation}
where $i=1,\ldots,N$. N is the number of orbiting objects, $\gamma$
a very small number and $T$ the resonance period. The simplest case
of resonance is when two orbital periods (e.g. $P_{1}$ and $P_{2}$)
have an integer ratio: $P_{1}/P_{2}=n$, where n is the integer 1,
2 or 3 etc. 

A linear resonance is also stable if its resonance order is zero,
that is if $\sum a_{i}=0$. Stable resonances are independent on the
selection of the rotating reference system. In fact, relative to any
observer moving with any period M with regard to stars, each orbital
body would rotate with a frequency $f_{iM}=1/T_{i}-1/M$. It is easy
to demonstrate that for stable resonances it holds: $T^{-1}=\sum a_{i}/T_{i}=\sum a_{i}f_{iM}$.
Therefore, stable resonances significantly characterize the physical
properties of an orbital system.

The simplest cases of stable resonances are the synodical periods
between two orbiting objects whose frequency is given by $f_{12}=\left|1/P_{1}-1/P_{2}\right|$.
For example: the synodic or conjunction period between Jupiter ($P_{J}=4332.589$
days) and Saturn ($P_{S}=10759.22$ days) is $P_{JS}=(P_{J}^{-1}-P_{S}^{-1})^{-1}=7253.455$
days $=19.859$ years; the synodic or conjunction period between Uranus
($P_{U}=30685.4$ days) and Neptune ($P_{N}=60189.0$ days) is $P_{UN}=(P_{U}^{-1}-P_{N}^{-1})^{-1}=62599.94$
days $=171.393$ years. The orbital periods are from NASA (http://nssdc.gsfc.nasa.gov/planetary/factsheet/);
we use the conversion: 1 year = 365.2425 days.

There are very few stable orbital resonances and if the coefficients
$a_{i}$, which must be small, are chosen between -3 and 3 only one
resonance falls within the Hallstatt time scale of 2100-2500 year
and, in general, for periods larger than 200 years. This is a combination
of the orbital periods of Jupiter, Saturn, Uranus and Neptune such
that:

\begin{equation}
f_{JSUN}=\frac{1}{P_{j}}-\frac{3}{P_{S}}+\frac{1}{P_{U}}+\frac{1}{P_{N}}.
\end{equation}
The period of such a resonance is 

\begin{equation}
P_{JSUN}=\frac{1}{f_{JSUN}}=846471.447\:d=2317.56\:yr.
\end{equation}
Since the above resonance is stable, the same period can be determined
by any observer moving with any period M with regard to stars. The
physical meaning of the above resonance will be demonstrated in the
following sections. Herein we stress that this resonance involves
a combination of all four Jovian planets. We also notice that such
a resonance nearly corresponds to about 116.5 revolutions of the conjunction
period of Jupiter and Saturn ($116.5*19.859=2313.6$ yr), and 13.5
revolutions of the conjunction period of Uranus and Neptune ($13.5*171.393=2313.8$
yr). Thus, every about 2313.7 years there exists a repeating pattern
involving conjunctions and oppositions among the four Jovian planets
of the solar system whose gravitational effect is revealed in the
following sections. 

\begin{table}
\centering{}\begin{tabular}{|c|c|c|c|c||c|c|c|c|c|c|} \hline  
$a_{Jup}$	&	$a_{Sat}$	&	$a_{Ura}$	&	$a_{Nep}$	& T (yr) & $a_{Jup}$	&	$a_{Sat}$	&	$a_{Ura}$	&	$a_{Nep}$	& T (yr)	
\tabularnewline \hline \hline 
3	&	-1	&	-2	&	0	&	5.12	&	2	&	-2	&	-2	&	2	&	11.23	\tabularnewline \hline 2	&	2	&	-3	&	-1	&	5.14	&	1	&	1	&	-3	&	1	&	11.29	\tabularnewline \hline 3	&	-2	&	2	&	-3	&	5.25	&	2	&	-3	&	2	&	-1	&	11.83	\tabularnewline \hline 3	&	-1	&	-3	&	1	&	5.28	&	1	&	0	&	1	&	-2	&	11.90	\tabularnewline \hline 3	&	-2	&	1	&	-2	&	5.41	&	0	&	3	&	0	&	-3	&	11.96	\tabularnewline \hline 2	&	1	&	0	&	-3	&	5.42	&	2	&	-2	&	-3	&	3	&	12.02	\tabularnewline \hline 3	&	-2	&	0	&	-1	&	5.59	&	2	&	-3	&	1	&	0	&	12.71	\tabularnewline \hline 2	&	1	&	-1	&	-2	&	5.60	&	1	&	0	&	0	&	-1	&	12.78	\tabularnewline \hline 3	&	-2	&	-1	&	0	&	5.78	&	0	&	3	&	-1	&	-2	&	12.85	\tabularnewline \hline 2	&	1	&	-2	&	-1	&	5.79	&	2	&	-3	&	0	&	1	&	13.73	\tabularnewline \hline 3	&	-3	&	3	&	-3	&	5.93	&	1	&	0	&	-1	&	0	&	13.81	\tabularnewline \hline 3	&	-2	&	-2	&	1	&	5.98	&	0	&	3	&	-2	&	-1	&	13.90	\tabularnewline \hline 2	&	1	&	-3	&	0	&	5.99	&	1	&	-1	&	3	&	-3	&	14.74	\tabularnewline \hline 3	&	-3	&	2	&	-2	&	6.15	&	2	&	-3	&	-1	&	2	&	14.93	\tabularnewline \hline 2	&	0	&	1	&	-3	&	6.16	&	1	&	0	&	-2	&	1	&	15.02	\tabularnewline \hline 3	&	-2	&	-3	&	2	&	6.19	&	0	&	3	&	-3	&	0	&	15.12	\tabularnewline \hline 3	&	-3	&	1	&	-1	&	6.37	&	1	&	-1	&	2	&	-2	&	16.12	\tabularnewline \hline 2	&	0	&	0	&	-2	&	6.39	&	0	&	2	&	1	&	-3	&	16.24	\tabularnewline \hline 1	&	3	&	-1	&	-3	&	6.41	&	2	&	-3	&	-2	&	3	&	16.35	\tabularnewline \hline 3	&	-3	&	0	&	0	&	6.62	&	1	&	0	&	-3	&	2	&	16.47	\tabularnewline \hline 2	&	0	&	-1	&	-1	&	6.64	&	1	&	-1	&	1	&	-1	&	17.80	\tabularnewline \hline 1	&	3	&	-2	&	-2	&	6.66	&	0	&	2	&	0	&	-2	&	17.93	\tabularnewline \hline 3	&	-3	&	-1	&	1	&	6.89	&	1	&	-1	&	0	&	0	&	19.86	\tabularnewline \hline 2	&	0	&	-2	&	0	&	6.91	&	0	&	2	&	-1	&	-1	&	20.03	\tabularnewline \hline 1	&	3	&	-3	&	-1	&	6.93	&	1	&	-1	&	-1	&	1	&	22.46	\tabularnewline \hline 2	&	-1	&	2	&	-3	&	7.13	&	0	&	2	&	-2	&	0	&	22.68	\tabularnewline \hline 3	&	-3	&	-2	&	2	&	7.17	&	1	&	-2	&	3	&	-2	&	25.01	\tabularnewline \hline 2	&	0	&	-3	&	1	&	7.20	&	0	&	1	&	2	&	-3	&	25.29	\tabularnewline \hline 2	&	-1	&	1	&	-2	&	7.44	&	1	&	-1	&	-2	&	2	&	25.85	\tabularnewline \hline 1	&	2	&	0	&	-3	&	7.46	&	0	&	2	&	-3	&	1	&	26.14	\tabularnewline \hline 3	&	-3	&	-3	&	3	&	7.49	&	1	&	-2	&	2	&	-1	&	29.29	\tabularnewline \hline 2	&	-1	&	0	&	-1	&	7.78	&	0	&	1	&	1	&	-2	&	29.66	\tabularnewline \hline 1	&	2	&	-1	&	-2	&	7.80	&	1	&	-1	&	-3	&	3	&	30.44	\tabularnewline \hline 2	&	-1	&	-1	&	0	&	8.15	&	1	&	-2	&	1	&	0	&	35.32	\tabularnewline \hline 1	&	2	&	-2	&	-1	&	8.18	&	0	&	1	&	0	&	-1	&	35.87	\tabularnewline \hline 2	&	-2	&	3	&	-3	&	8.46	&	1	&	-2	&	0	&	1	&	44.49	\tabularnewline \hline 2	&	-1	&	-2	&	1	&	8.55	&	0	&	1	&	-1	&	0	&	45.36	\tabularnewline \hline 1	&	2	&	-3	&	0	&	8.58	&	0	&	0	&	3	&	-3	&	57.13	\tabularnewline \hline 2	&	-2	&	2	&	-2	&	8.90	&	1	&	-2	&	-1	&	2	&	60.09	\tabularnewline \hline 1	&	1	&	1	&	-3	&	8.93	&	0	&	1	&	-2	&	1	&	61.69	\tabularnewline \hline 2	&	-1	&	-3	&	2	&	9.00	&	1	&	-3	&	3	&	-1	&	82.64	\tabularnewline \hline 2	&	-2	&	1	&	-1	&	9.39	&	0	&	0	&	2	&	-2	&	85.70	\tabularnewline \hline 1	&	1	&	0	&	-2	&	9.42	&	-1	&	3	&	1	&	-3	&	88.99	\tabularnewline \hline 2	&	-2	&	0	&	0	&	9.93	&	1	&	-2	&	-2	&	3	&	92.54	\tabularnewline \hline 1	&	1	&	-1	&	-1	&	9.97	&	0	&	1	&	-3	&	2	&	96.39	\tabularnewline \hline 2	&	-2	&	-1	&	1	&	10.54	&	1	&	-3	&	2	&	0	&	159.59	\tabularnewline \hline 1	&	1	&	-2	&	0	&	10.59	&	0	&	0	&	1	&	-1	&	171.39	\tabularnewline \hline 2	&	-3	&	3	&	-2	&	11.07	&	-1	&	3	&	0	&	-2	&	185.08	\tabularnewline \hline 1	&	0	&	2	&	-3	&	11.12	&	1	&	-3	&	1	&	1	&	2317.56	\tabularnewline \hline 
 \end{tabular}\caption{Stable resonances associated to the Jupiter-Saturn-Uranus-Neptune
system. The coefficients $a_{i}$ of Eq. \ref{eq:2} are made to vary
between -3 and 3. See also Figure 4B.}
\end{table}

Additional resonances can be calculated by making the coefficients
$a_{i}$ to vary within a larger range. However, even if this range
is chosen to be between -10 and 10, the only stable resonance periods
found for periods larger than 1000 years are at 1158.78 yr (resonance
2:-6:2:2), 1159.30 yr (resonance -1:2:4:-5), 2317.56 yr (resonance
1:-3:1:1), 2319.62 yr (resonance -2:5:3:-6) and 2,607,251.87 yr (resonance
3:-8:-2:7). Thus, the period of 2317-2320 years represents a very
important and unique stable resonance of the solar system. We note
that \citet{Humlum} found a $1139\pm160$ yr oscillation in the detrended
GISP2 surface temperature series during the last 4000 years and \citet{Davis}
found a spectral peak between 950 yr and 1113 yr, which may be coherent
to the above millennial stable resonances although for the Eddy cycle
there could be alternative explanations \citep[e.g.: ][]{Scafetta2012a,Scafetta2014}.

Two other important stable resonances that we will meet in the next
sections are:

\begin{equation}
f_{JSU}=\frac{1}{P_{J}}-\frac{3}{P_{S}}+\frac{2}{P_{U}},
\end{equation}
which gives $P_{JSU}=159.59$ yr and

\begin{equation}
f_{JSN}=-\frac{1}{P_{J}}+\frac{3}{P_{S}}-\frac{2}{P_{N}},
\end{equation}
which gives $P_{JSN}=185.08$ yr.

We will demonstrate that, using opportune astronomical observables,
the 2318-year resonance appears like a modulation of these faster
resonance oscillations together with that of the Uranus-Neptune synodic
stable resonance $P_{UN}=171.393$ yr. The three 159-171-185 yr astronomical
resonances are very important also for supporting the main hypothesis
of our paper, namely to interpret astronomically the origin of the
Hallstatt oscillation observed in radionucleotide and climate records.
In fact, these resonances, in particular $P_{UN}$, have already been
found to characterize such geophysical records used to reconstruct
also cosmic ray flux and solar activity throughout the Holocene \citep[cf.: ][]{McCracken2014,Sharp}
and also with aurora records available since the 16th century \citep{ScafettaWillson2013a}. 

Table 1 reports a list of stable resonances for periods larger than
5 years associated to the Jupiter-Saturn-Uranus-Neptune system where
the coefficients $a_{i}$ are made to vary between -3 and 3. These
resonances are clustered around specific frequencies. In particular,
note the resonance clusters at 44-46 years, 57-62 years and 82-97
years that are found in solar and aurora activity \citep[e.g.: ][]{McCracken2001,Ogurtsov,ScafettaWillson2013a,Scafetta2014,Vaquero}
and also in climate records \citep[e.g.: ][and many others]{Czymzik,Hoyt},
e.g. a quasi 60-year cycle is very important in climate \citep[e.g.:][and many others]{Gervais,Loehle,Manzi,Mazzarella,Scafetta2010,Scafetta2014c,Wyatt}.
The 82-97 year period is known as the Gleissberg cycle.

In the next section we will construct physical observables that better
reveal the above Jupiter-Saturn-Uranus-Neptune resonances. From a
purely spectral point of view, it may be pointed out that many functions
of the orbits of the planets (e.g. total angular momentum of the planets,
speed and position of the PMC relative to the Sun, etc) are expected
to present numerous common spectral peaks simply because the harmonic
input would be the same. We will use the eccentricity function of
the Sun-PMC orbit and other specific orbital proxies because these
proxies suggest a possible physical mechanism, as we will discuss
in the next sections. 

\section{The eccentricity vector}

Let us fully derive the instantaneous eccentricity function of the
orbit of a generic planet orbiting the Sun \citep[cf: ][]{Mungan}.
In classical celestial mechanics a Keplerian orbit is defined as the
motion of an object orbiting another (e.g. a planet orbiting its star)
under Newton's force of gravity:

\begin{equation}
m\boldsymbol{a}=-\frac{GMm}{r^{2}}\boldsymbol{\hat{r}},\label{eq:1}
\end{equation}
Because $\boldsymbol{r}=r\hat{\boldsymbol{r}}$ and the angular momentum
$\boldsymbol{L}=m\boldsymbol{r}\boldsymbol{\times v}$ is constant,
Eq. \ref{eq:1} can be easily rewritten as

\begin{equation}
\frac{d}{dt}(\boldsymbol{v}\boldsymbol{\times L})=GMm\frac{d\boldsymbol{\hat{r}}}{dt},\label{eq:3}
\end{equation}
where $\boldsymbol{v}=d\boldsymbol{r}/dt$ is the velocity of the
orbiting body. By integrating Eq. \ref{eq:3}, we obtain

\begin{equation}
\boldsymbol{v}\times\boldsymbol{L}=GMm(\boldsymbol{\hat{r}}+\boldsymbol{e}),\label{eq:4}
\end{equation}
where $\boldsymbol{e}$ is an integration constant vector. After a
simple vector algebra we obtain 

\begin{equation}
\boldsymbol{e}=\frac{\boldsymbol{v}\times(\boldsymbol{r}\times\boldsymbol{v})}{GM}-\boldsymbol{\hat{r}}.\label{eq:5}
\end{equation}
To understand the physical meaning of the vector $\boldsymbol{e}$,
we take the dot product of the position vector $\boldsymbol{r}$ with
Eq. \ref{eq:4} to obtain

\begin{equation}
\boldsymbol{r}\cdot(\boldsymbol{v}\boldsymbol{\times L})=GMm\boldsymbol{r\cdot}(\boldsymbol{\hat{r}}+\boldsymbol{e})\label{eq:6}
\end{equation}
that becomes

\begin{equation}
m(\boldsymbol{r}\times\boldsymbol{v})\boldsymbol{\cdot L}=L^{2}=GMm^{2}(1+e\cos\theta)r.\label{eq:7}
\end{equation}
If $c=L^{2}/(GMm^{2})$ and the vector $\boldsymbol{e}$ is chosen
to point towards the periapsis of the orbit, in the traditional $r$-$\theta$
polar coordinates, Eq. \ref{eq:7} is equivalent to the traditional
Keplerian orbital equation:

\begin{equation}
r(\theta)=\frac{c}{1+e\cos\theta},\label{eq:2-1}
\end{equation}
where $c$ is a constant called the semi-latus rectum of the curve
and $e$ is the eccentricity of the orbit. 

For a circular orbit $e=0$; for an elliptical orbit $0<e<1$ and
$\theta=0$ and $\theta=\pi$ indicate the position of the perihelion
and aphelion, respectively; for a parabolic trajectory $e=1$; and
for a hyperbolic trajectory $e>1$. In the case of a simple two-body
system, without any form of dissipation or perturbation, celestial
mechanics predicts that the eccentricity $e$ of an orbit is constant.
Thus, Eq. \ref{eq:5} defines the eccentricity vector, whose scalar
is the eccentricity of the Keplerian orbit.

Using simple vector algebra, $\boldsymbol{v\times}(\boldsymbol{r\times}\boldsymbol{v})=(\boldsymbol{v\cdot}\boldsymbol{v})\boldsymbol{r}-(\boldsymbol{r}\boldsymbol{\cdot v})\boldsymbol{v}$,
Eq. \ref{eq:5} can be rewritten in a more friendly way, and the instantaneous
eccentricity of the trajectory of each planet of the solar system
can be defined as

\begin{equation}
e=\left|\left(\frac{v^{2}}{\mu}-\frac{1}{r}\right)\boldsymbol{r}-\frac{\boldsymbol{r}\cdot\boldsymbol{v}}{\mu}\boldsymbol{v}\right|,\label{eq:8-1}
\end{equation}
where $\mu=GM_{sun}=2.959122082855911\cdot10^{-4}$ $AU^{3}/d^{2}$
is the standard gravitational parameter for the Sun as used in the
adopted ephemeris files \citep[Table 8]{Folkner}. However, Eq. \ref{eq:1}
works if $M\gg m$. In real cases, the mass $m$ on the left side
of \ref{eq:1} must be substituted with the reduced mass, $M_{sun}m_{planet}/(M_{sun}+m_{planet})$,
which yields again to Eq. \ref{eq:8-1} with the following correction
$\mu=G(M_{sun}+m_{planet})$, as we will use in the following section.

\section{Definition of the planetary mass center relative to the Sun}

The wobbling of the Sun occurs mostly close to the ecliptic orbital
plane and it is a real feature of the solar system relative from the
outer deep space from where the cosmic ray flux comes. We hypothesized
that the rhythmic contraction and expansion of the solar system disk
associated to its inner wobbling could modulate the incoming cosmic
ray flux reaching the Earth and/or alter the physical properties of
the heliosphere modulating the solar wind and interplanetary/cosmic
dust concentration. This dynamics can be represented by the movement
of the planetary mass center (PMC) relative to the Sun. This orbit
is deduced in the following way using programs that implement the
ephemeris files DE431/DE432 prepared by the NASA Solar System Dynamics
Group of the Jet Propulsion Laboratory (ftp://ssd.jpl.nasa.gov/pub/eph/planets/ascii)
\citep{Folkner,Folkner-1}.

\citet{Folkner} reports that the orientation of the DE431/DE432 ephemeris
is tied to the International Celestial Reference Frame with an accuracy
of 0.0002'': for the inner planets the orbital accuracy is of the
order of a few hundred meters, for Jupiter and Saturn the orbital
accuracy is of tens of kilometers and for Uranus, Neptune, and Pluto
the orbital accuracy worsen up to several thousand kilometers. This
means that the orbital parameters have at least a 9 to 7 digit precision
from the inner planets up to Pluto, respectively. Thus, the ephemeris
error-bars alone are not expected to provide false evidences for major
cycles, also because several of the observed spectral peaks of the
adopted astronomical observables can be easily recognized as stable
orbital resonances or orbital periods, as shown below.

Let $\boldsymbol{r}_{p}$ be the vector position of the center of
mass of all objects of the solar system, Sun excluded, relative to
the barycenter; let $\boldsymbol{r}_{s}$ be the vector position of
the Sun relative to barycenter; let $M_{p}$ be the total mass of
all objects of the solar system excluded the Sun, that is the sum
of the 352 masses (planets + asteroids) taken into account by the
DE432 JPL ephemeris file; let $M_{s}$ be the mass of the Sun. These
masses are deduced from the parameters of the header file of the NASA
JPL DE432 ephemeris file.

Relative to the barycenter, the position vectors and the relative
velocities are balanced, that is, observing that $M_{p}\boldsymbol{r}_{p}=\sum_{i}m_{i}\boldsymbol{r}_{i}$
and $M_{p}\boldsymbol{v}_{p}=\sum_{i}m_{i}\boldsymbol{v}_{i}$ where
the index i refers to each planetary or asteroid object of the solar
system, the following equations are fulfilled:

\begin{equation}
M_{s}\boldsymbol{r}_{s}+M_{p}\boldsymbol{r}_{p}=0
\end{equation}

\begin{equation}
M_{s}\boldsymbol{v}_{s}+M_{p}\boldsymbol{v}_{p}=0.
\end{equation}
Let $\boldsymbol{r}$ and $\boldsymbol{v}$ be the position and velocity
vector, respectively, of the PMC relative to the the Sun. Thus, we
have:

\begin{equation}
\boldsymbol{r}=-\boldsymbol{r}_{s}+\boldsymbol{r}_{p}=\frac{M_{s}+M_{p}}{M_{p}}(-\boldsymbol{r}_{s})\label{33}
\end{equation}

\begin{equation}
\boldsymbol{v}=-\boldsymbol{v}_{s}+\boldsymbol{v}_{p}=\frac{M_{s}+M_{p}}{M_{p}}(-\boldsymbol{v}_{s})\label{eq:44}
\end{equation}
Figure 2 shows a section of the orbit of PMC from 1950 to 2050 where
it is seen that the PMC point wobbles around the Sun within an orbit
of about 7 AU radius.

Although Eq. \ref{eq:8-1} can be rigorously applied only to a Keplerian
orbit, and is useful to determine for example the small fluctuations
of the orbits of the planets of the solar system, we can assume that
the PMC, which does not follow a Keplerian orbit, at each instant
represents a given planet P that is orbiting the Sun at that specific
position $\boldsymbol{r}$ and with that specific velocity $\boldsymbol{v}$
estimated in Eqs. \ref{33} and \ref{eq:44}. Then, we define the
instantaneous eccentricity of the orbit of PMC as the eccentricity
of the hypothetical orbit of the planet P evolving in time using Eq.
\ref{eq:8-1}, the $\boldsymbol{r}$ and $\boldsymbol{v}$ vectors
estimated in Eqs. \ref{33} and \ref{eq:44} and $\mu=G(M_{s}+M_{CMP})=2.963092749817812\cdot10^{-4}$
$AU^{3}/d^{2}$, which takes into account all masses of the solar
system used in the ephemeris files DE431/DE432.

\begin{figure}[!t]
\centering{}\includegraphics[width=1\textwidth]{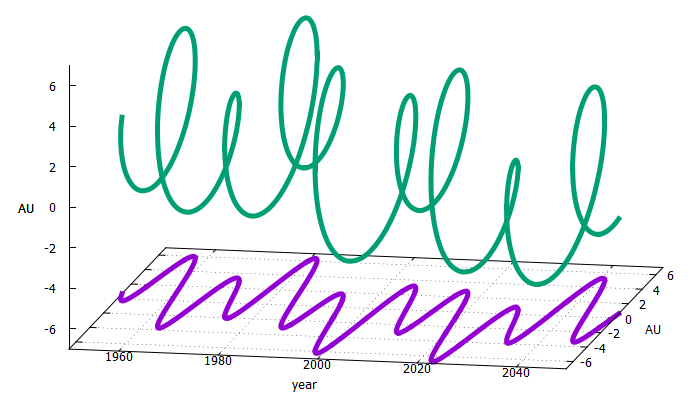}\caption{Motion of the PMC relative to the Sun from 1950 to 2050. }
\end{figure}

\begin{figure}[!t]
\centering{}\includegraphics[width=1\textwidth]{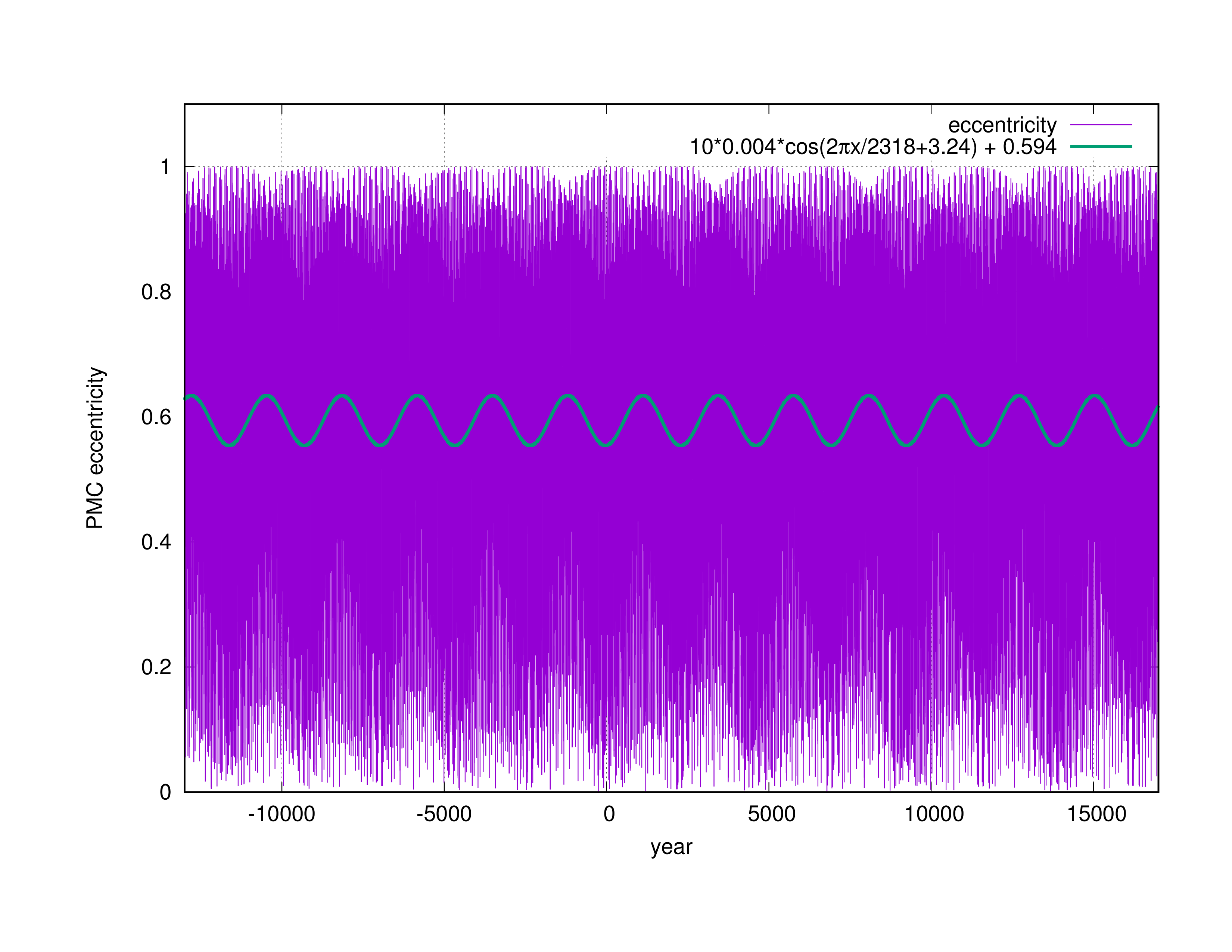}\caption{Variation of the eccentricity (Eq. \ref{eq:8-1}) of the PMC relative
to the Sun. The latter is fit with a sinusoidal function (green) whose
amplitude has been magnified by 10 for visual convenience.}
\end{figure}

\section{Analysis of the ``eccentricity'' variation of the Sun-PMC orbit}

Figure 3 shows the record of the eccentricity $e$ of the Sun-PMC
orbit calculated by Eq. \ref{eq:8-1} from 13,000 B. C. to 17,000
A. D. sampled every 30 days. Figure 4A shows its periodogram and Figure
4B compares it with the periods of the stable planetary resonances
reported in Table 1. A clear correspondence is found between all spectral
peaks of the eccentricity function of the Sun-PMC orbit and the stable
resonances generated by Jupiter, Saturn, Uranus and Neptune. 

As expected, the instantaneous eccentricity of the Sun-PMC orbit varies
greatly from a trajectory nearly circular ($e\approx0$) to one nearly
parabolic ($e\approx1$). Large oscillations are observed, in particular
at the 19.86 yr synodic period between Jupiter and Saturn. Other strong
oscillations close to the known planetary orbital periods of the four
Jovian planets are observed: Jupiter, 11.86 yr; Saturn, 29.46 yr;
Uranus, 84.01 yr; Neptune, 164.8 yr. In addition, several other synodic
periods among planets are observed as well: Earth-Jupiter, 1.092 yr;
Jupiter-Uranus, 13.8 yr; Jupiter-Neptune, 12.78 yr; Saturn-Uranus,
45.4 yr; Saturn-Neptune, 38.9 yr; Uranus-Neptune, 172 yr. Also the
Jupiter-Saturn trigon synodic period, 57-61 yr, is well observed.
All these known oscillations were well expected. 

For what concerns this study, Figures 3 and 4 also demonstrate that
the chosen eccentricity function presents a major oscillation at about
2100-2500 yr period that could not be immediately derived from the
individual planetary orbital periods. The statistical error of the
periodogram associated to a spectral peak period is $\nabla p=\pm p^{2}/2L$,
where $L=30000$ yr is the length of the record analyzed. The observed
periodogram peak period is at $p=2318\pm90$ yr. Thus, it is evidently
due to the Jupiter-Saturn-Uranus-Neptune resonance discussed in section
3. This periodicity is perfectly coherent with the Hallstatt oscillation
found in the radionucleotide records as that shown in Figures 1: see
Figure 10. The periodogram depicted in Figure 4 stresses that the
2318 yr period peak is the most relevant within the spectral range
between 200 and 10000 yr periods indicating that this oscillation
dominates this time scale, as also found for the stable resonances
reported in Table 1.

Figure 5 compares the cosine curves used to fit both the radionucleotide
record depicted in Figure 1B, and the eccentricity function depicted
in Figure 3. The periods are the same, within their error of measure,
and the phases are $\phi\approx1.82$ and $\phi\approx3.24$, respectively.
Thus, as Figure 5 (upper panel) shows, the two harmonics are shifted
by almost $\pi/2$ or about 525 years. This means that the Hallstatt
oscillation of the radionucleotide record is nearly proportional to
the integration or to the negative of the derivative of the eccentricity
record of the Sun-PMC orbit. 

\begin{figure}[!t]
\centering{}\includegraphics[width=1\textwidth]{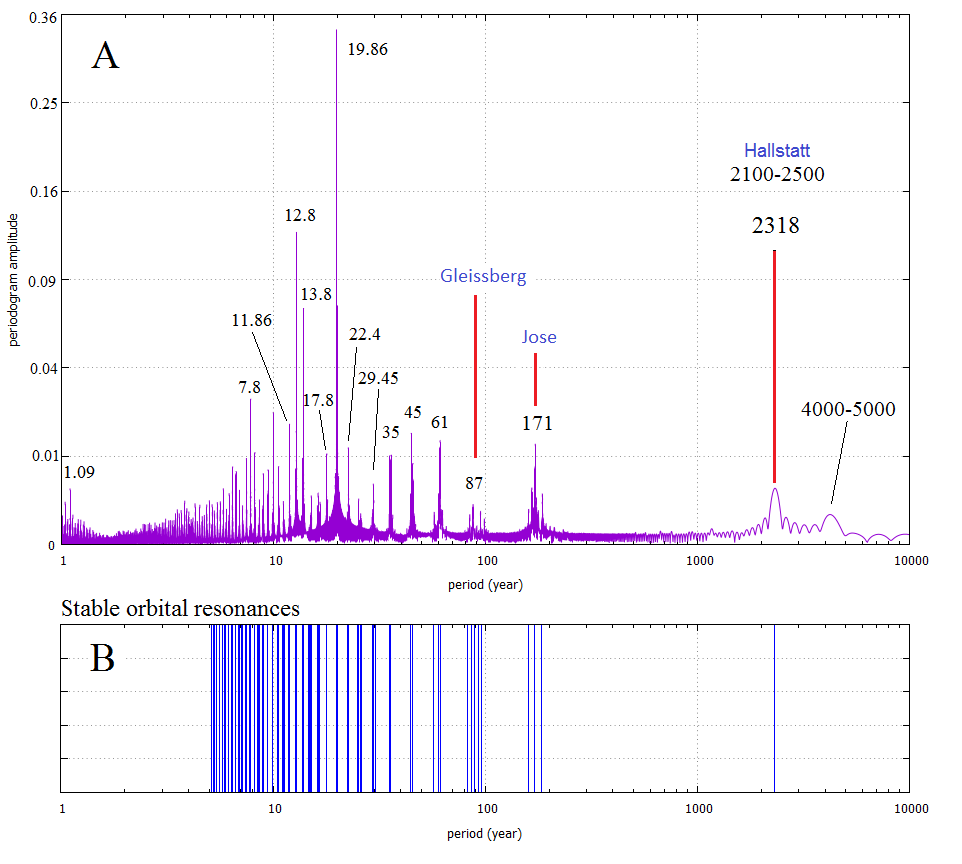}\caption{{[}A{]} Peridogram of the eccentricity record of the Sun-PMC orbit
depicted in Figure 3. The spectral peak corresponding at the Hallstatt
period at 2318 yr is well visible. {[}B{]} The blue bars represent
the stable planetary resonances of the solar system generated by Jupiter,
Saturn, Uranus and Neptune for periods larger than 5 years: see also
Table 1. Note the accurate correspondence between these resonances
and the spectral peaks depicted in {[}A{]}.}
\end{figure}

\begin{figure}[!t]
\centering{}\includegraphics[width=1\textwidth]{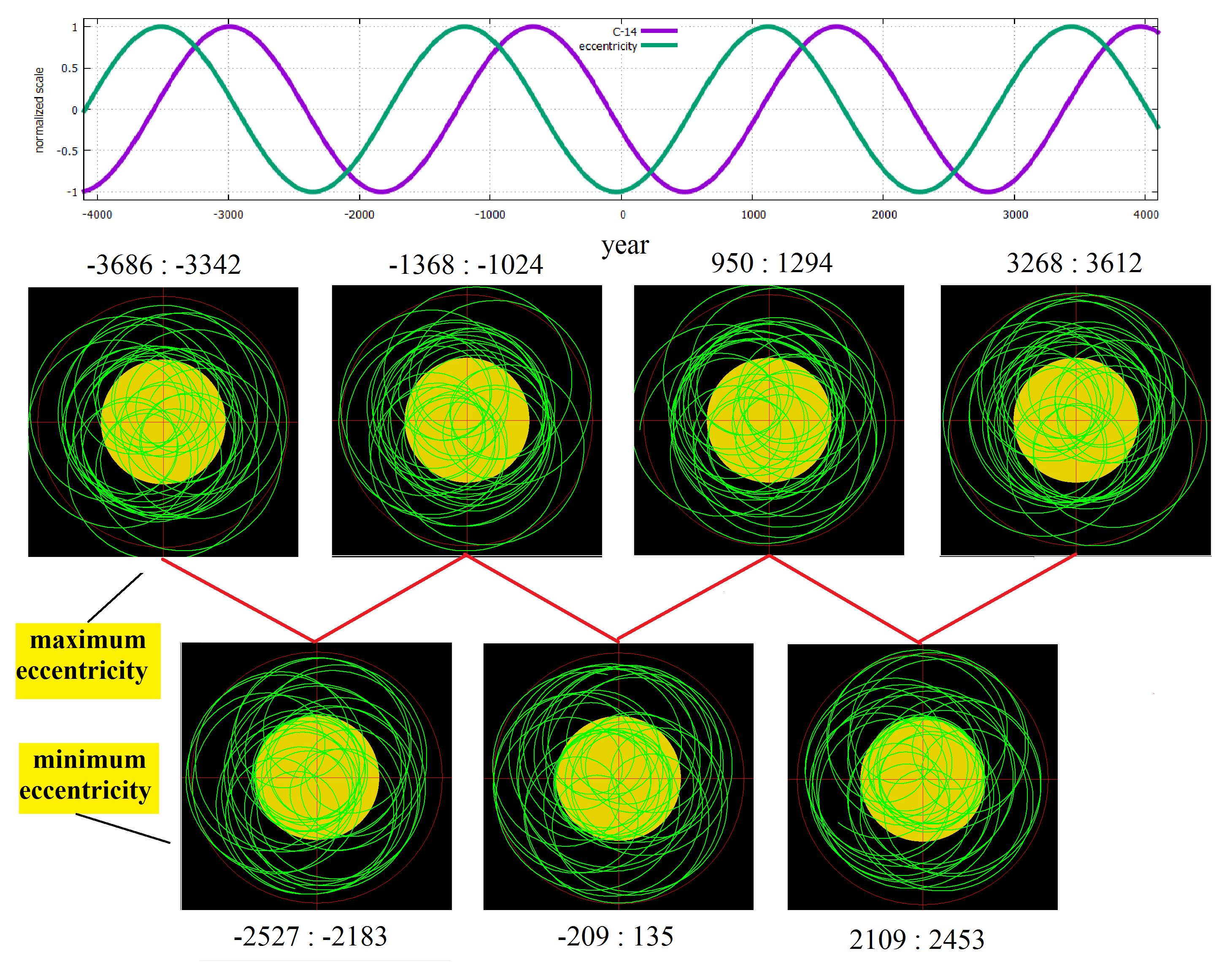}\caption{(Top) 2300-2400 yr harmonics referring to the $\varDelta^{14}C$ (\textperthousand )
record depicted in Figure 1 and the eccentricity record depicted in
Figure 3. Note the $\pi/2$ phase shift. (Bottom) Trajectory of the
PMC relative to the Sun during intervals of 344 years referring to
periods of the eccentricity maxima (above, more disordered, open orbits)
and minima (below, more ordered, closed orbits). The yellow disk radius
is about 3.5 AU, while the red circle radius is about 7 AU. The transition
periods (red segments) from the eccentricity maxima to minima correspond
to maxima in radionucleotide production, while the transition periods
from the eccentricity minima to maxima correspond to minima in radionucleotide
production.}
\end{figure}

Figure 5 shows that, on the Hallstatt-cycle time scale, a larger production
of radionucleotide particles occurs while the Sun-PMC orbit evolves
from statistically more elliptical shapes ($e\approx0.598$) to statistically
more circular ones ($e\approx0.590$), that is while the system is
bursting outward. Analogously, a smaller production of radionucleotide
particles occurs while the Sun-PMC orbit evolves from statistically
more circular shapes ($e\approx0.590$) to statistically more elliptical
ones ($e\approx0.598$), that is while the system is bursting inward.

Figure 5 (lower panels) shows trajectories of the Sun-PMC orbits when
these are statistically more elliptical (upper list of lower panels,
the average eccentricity is $e\approx0.598$) and when these are statistically
more circular (lower list of lower panels, the average eccentricity
is $e\approx0.590$). The chosen time intervals were 344 year long
that is twice the 172 yr harmonic revealed in the periodogram of Figure
4. The upper list of these panels reveals that during these periods
the Sun-PMC orbits are skewed with large regions that are rarely visited
and the trajectory appears developing mostly within a 5 AU radius,
which is the orbit of Jupiter, but sometimes it also clearly exceeds
the 7 AU radius distance from the Sun (red curve). The lower list
of panels reveals that during these periods the Sun-PMC orbits are
more regular, more circular, symmetric and more uniformly cover all
areas within a 7 AU radius distance from the Sun. 

The dynamics observed in Figure 5 is also reminiscent at the larger
Hallstatt time scale of the trefoil ordered and disordered state of
the inertial motion of the Sun which is correlated to the gran maxima
and minima of solar activity, respectively, as suggested by \citet{Charvatova,Charv=0000E1tov=0000E1}
inspired by the 178.7 yr cycle found by \citet{Jose}. However, as
Figure 5 shows, here it is during the transition periods from an orbital
state to the other that correlates with periods of maximum or minimum
radionucleotide production. 

\begin{figure}[!t]
\centering{}\includegraphics[width=1\textwidth]{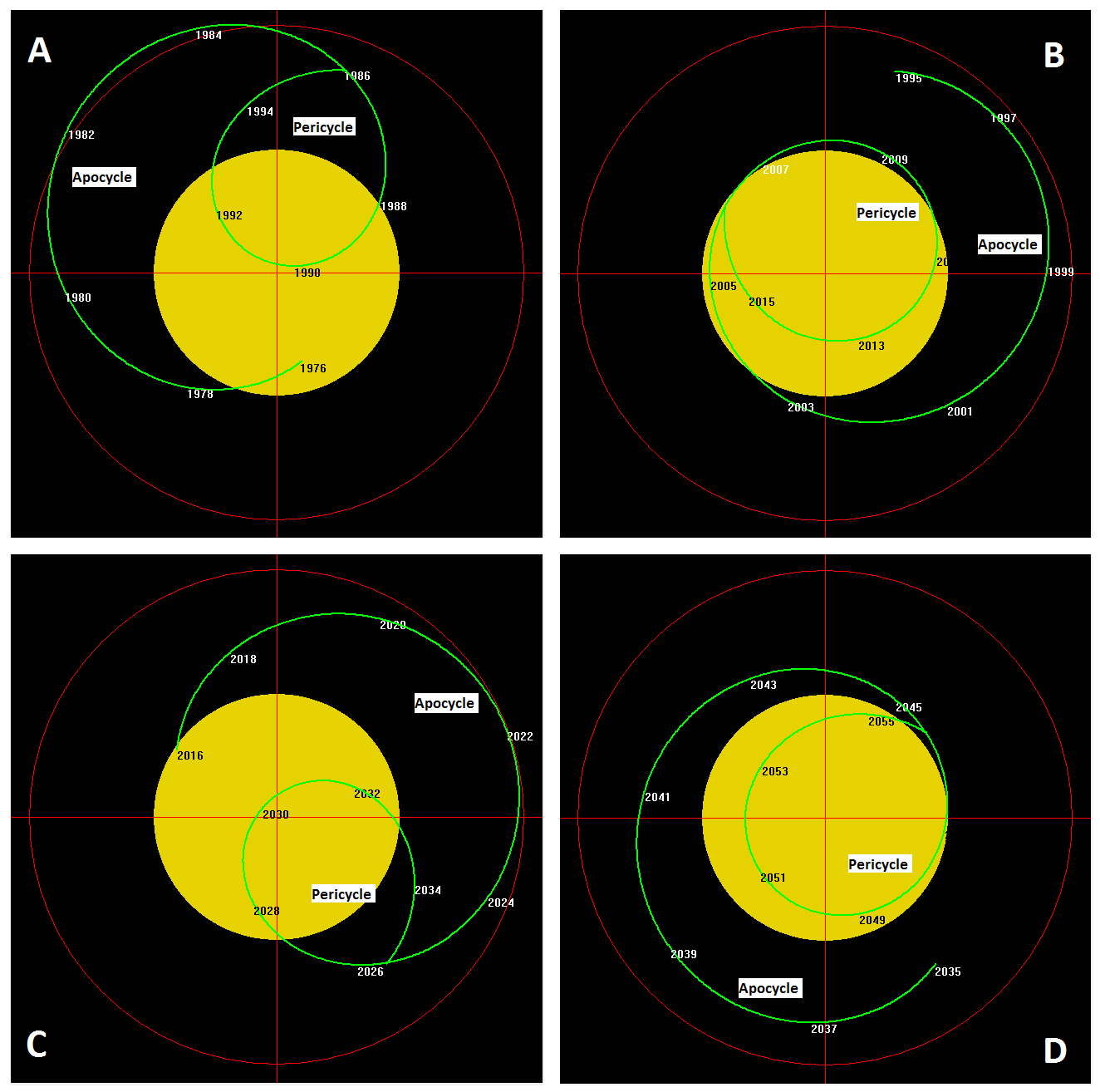}\caption{Motion of the PMC relative to the Sun from 10/6/1976 to 17/11/2055.
The diagrams depict four contiguous orbits made of one apocycle (external
larger orbit) and one pericycle (internal smaller orbit). (Solar Orbit
Simulator, http://arnholm.org/astro/sun/sc24/sim1/). The yellow disk
radius is about 3.5 AU, while the red circle radius is about 7 AU.}
\end{figure}

\begin{figure}[!t]
\centering{}\includegraphics[width=1\textwidth]{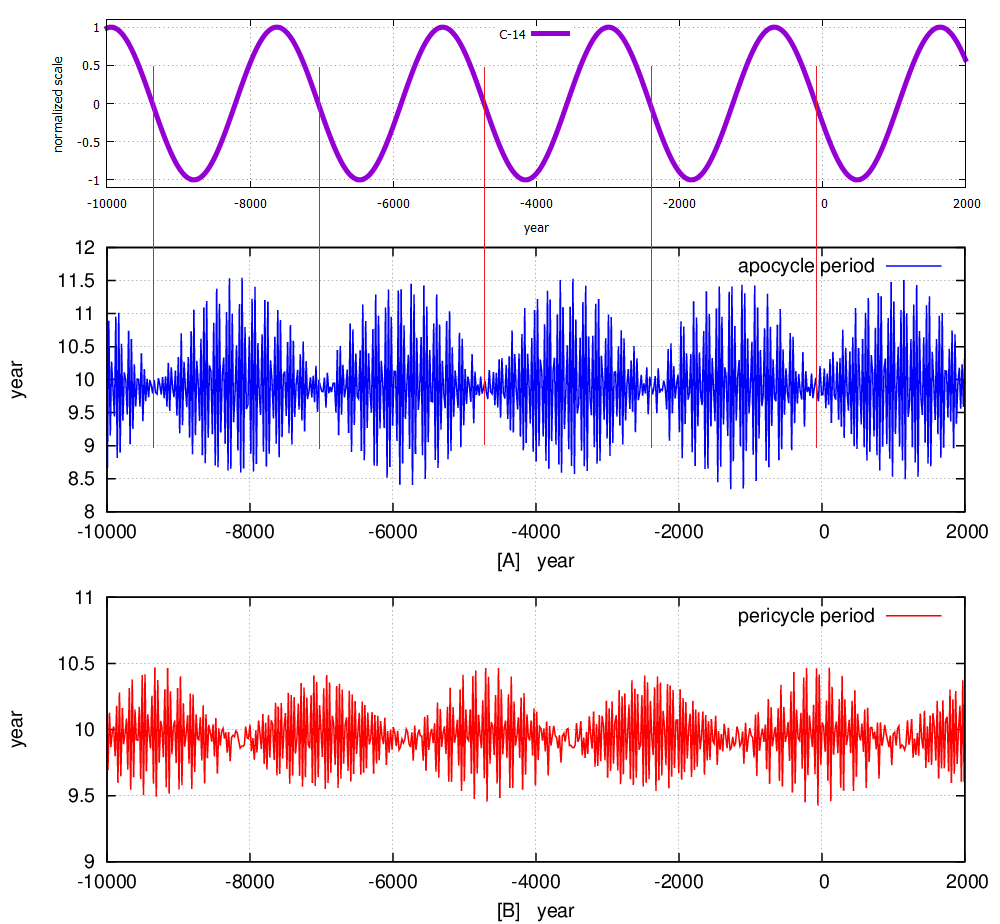}\caption{Upper panel: the Hallstatt oscillation found in $\varDelta^{14}C$
(\textperthousand ) record, as depicted in Figure 1B. Lower panels
show the time periods of the apocycles {[}A{]} and pericycles {[}B{]}
of the PMC relative to the Sun. The upper panel Hallstatt oscillation
is approximately in phase quadrature ($\phi=\pi/2$ and $\phi=-\pi/2$)
with the beat oscillation depicted in the lower panels as the red
vertical lines show. }
\end{figure}

\section{Analysis of the pericycle and apocycle orbital arcs}

The solar system pulses driven by the revolution of its planets around
the Sun and the major harmonic period of this dynamics within the
200-10,000 yr time scale is 2318 yr: this period perfectly corresponds
to the 2100-2500 yr Hallstatt oscillation. However, in the chosen
observable, the eccentricity variation of the Sun-PMC orbits, this
slow oscillation is relatively small: to make it visible in Figure
3 we needed to plot it magnified by 10. 

It is important to search for a more specific astronomical origin
for the Hallstatt-cycle that could stress the above dynamical characteristics
of the Sun-PMC orbit. The search for a more appropriate orbital proxy
is addressed in this section.

As evident in Figure 6, the Sun-PMC dynamics is characterized by a
series of unit cycles made of an apocycle, or external large orbit,
and a pericycle, or internal small orbit \citep[cf. ][]{Piovan}.
During each apocycle the PMC moves in an arc in which it reaches a
maximum speed and distance from the Sun. In the following pericycle
the PMC enters into a helical coil in which it reaches a minimum speed
and distance from the Sun before returning to a position very near
to the point where it has entered in the pericycle. This exit point
is the beginning of the following apocycle. Figure 6 shows 4 consecutive
full (apocycle plus pericycle) orbits from 10/6/1976 to 17/11/2055. 

\begin{figure}[!t]
\centering{}\includegraphics[width=1\textwidth]{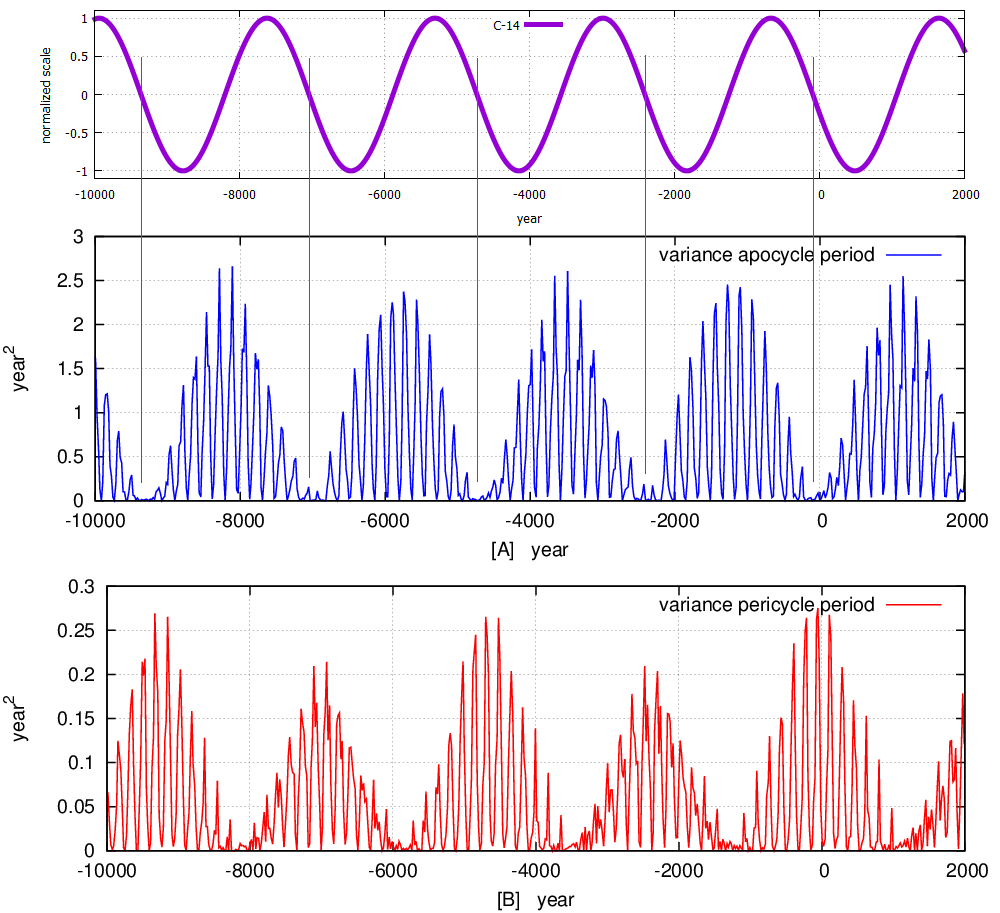}\caption{Upper panel: the Hallstatt oscillation found in $\varDelta^{14}C$
(\textperthousand ) record, as depicted in Figure 1B. Lower panels:
{[}A{]} and {[}B{]} are calculated from the records depicted in Figure
7A and B, respectively, as the square of their volatility from the
mean. The upper panel Hallstatt oscillation is approximately in phase
quadrature ($\phi=\pi/2$ and $\phi=-\pi/2$) with the oscillation
depicted in the lower panels as the red vertical lines show. }
\end{figure}

\begin{figure}[!t]
\centering{}\includegraphics[width=1\textwidth]{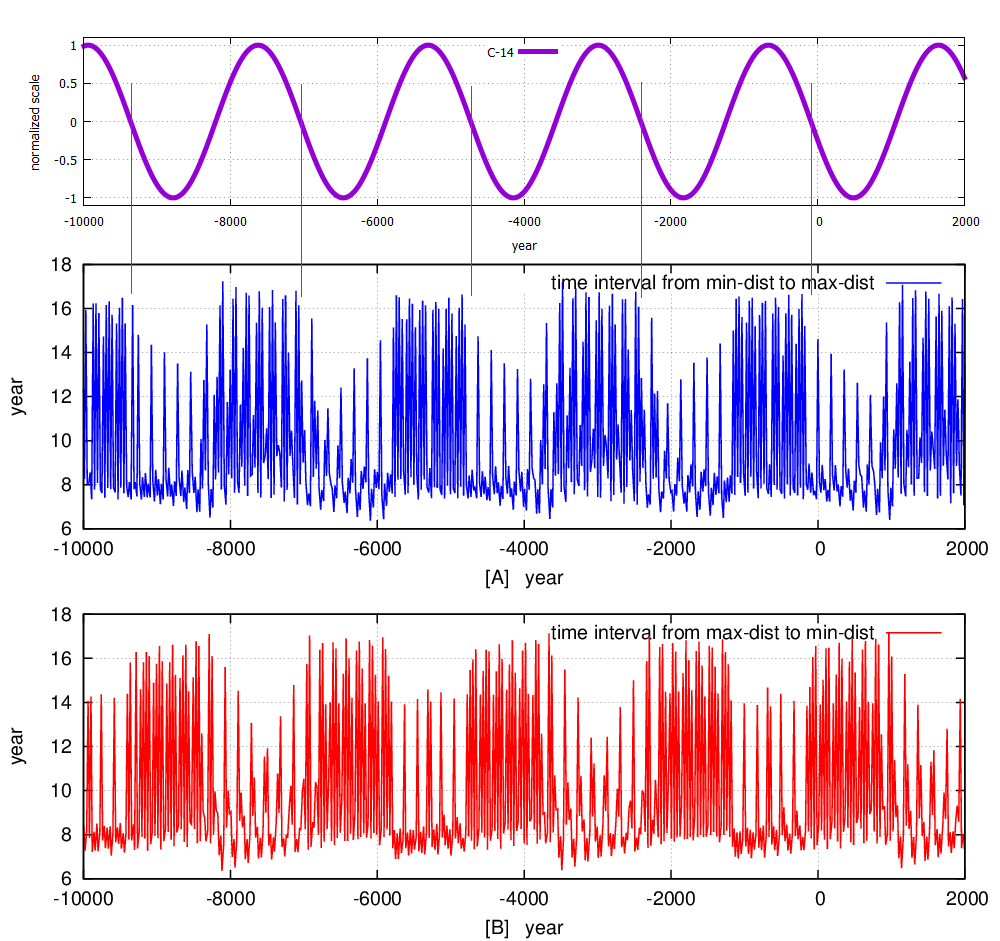}\caption{Upper panel: the Hallstatt oscillation found in $\varDelta^{14}C$
(\textperthousand ) record, as depicted in Figure 1B. Lower panels:
{[}A{]} Sequence of time intervals requested by the PMC to move from
its minimum to its maximum distance from the Sun; {[}B{]} Sequence
of time intervals requested by the PMC to move from its maximum to
its minimum distance from the Sun. The upper panel Hallstatt oscillation
is approximately in phase ($\phi=0$ and $\phi=-\pi$) with the oscillation
depicted in the lower panels as the red vertical lines show. }
\end{figure}

The orbit sections depicted in Figure 6 vary substantially in time.
Sometimes the apocycles are very different from the pericycle (as
in the figure). Other times they have similar amplitudes. This variation
is due to the relative position of the various planets, in particular
of the large Jovian ones. Let us investigate in details the dynamics
of these apocycles and pericycles.

Figure 7A shows the time periods of the subsequent apocycles, $P_{a}$,
while Figure 7B shows those of the following pericycles,$P_{p}$.
The apocycle periods average about $\mu_{a}=9.91$ yr and vary from
this mean up to a $\pm1.5$ years while the pericycle periods average
about $\mu_{p}=9.95$ yr and vary from their mean up to a $\pm0.5$
years. The sum of the two average periods is $19.86$ years that corresponds
to the conjunction period between Jupiter and Saturn. The upper panel
of Figure 7 shows an Hallstatt oscillation at 2318 yr period found
in $\varDelta^{14}C$ (\textperthousand ) record, as depicted in Figure
1B, to show that its phase is about $\phi=\pi/2$ and $\phi=-\pi/2$
with the beat oscillation depicted in the lower A and B panels, respectively. 

Figures 8A and 8B depict two records calculated from the periods depicted
in Figure 7A and B, respectively, as the square of the volatility
from their mean $\mu$, that is as: $(\Delta P_{a})^{2}=(P_{a}-\mu_{a})^{2}$
and $(\Delta P_{p})^{2}=(P_{p}-\mu_{p})^{2}$, respectively. This
operation was chosen to make even more evident the 2100-2500 year
oscillation present in these records. The upper panel of Figure 8
shows the Hallstatt oscillation found in $\varDelta^{14}C$ (\textperthousand )
record, as depicted in Figure 1B, to show that its phase is about
$\phi=\pi/2$ and $\phi=-\pi/2$ with the oscillation depicted in
the lower A and B panels, respectively. 

Finally, Figure 9A shows, for each full apocycle plus pericycle unit,
the times requested by the PMC to move from the minimum to the following
maximum distance from the Sun. Figure 9B, instead, shows for each
orbit the times requested by the PMC to move from the maximum to the
following minimum distance from the Sun. The upper panel of Figure
9 depicts the Hallstatt oscillation found in $\varDelta^{14}C$ (\textperthousand )
record, as in Figure 1B, to show that its phase is about $\phi=0$
and $\phi=\pi$ with the oscillation depicted in the lower A and B
panels, respectively. 

\begin{figure}[!t]
\centering{}\includegraphics[width=1\textwidth]{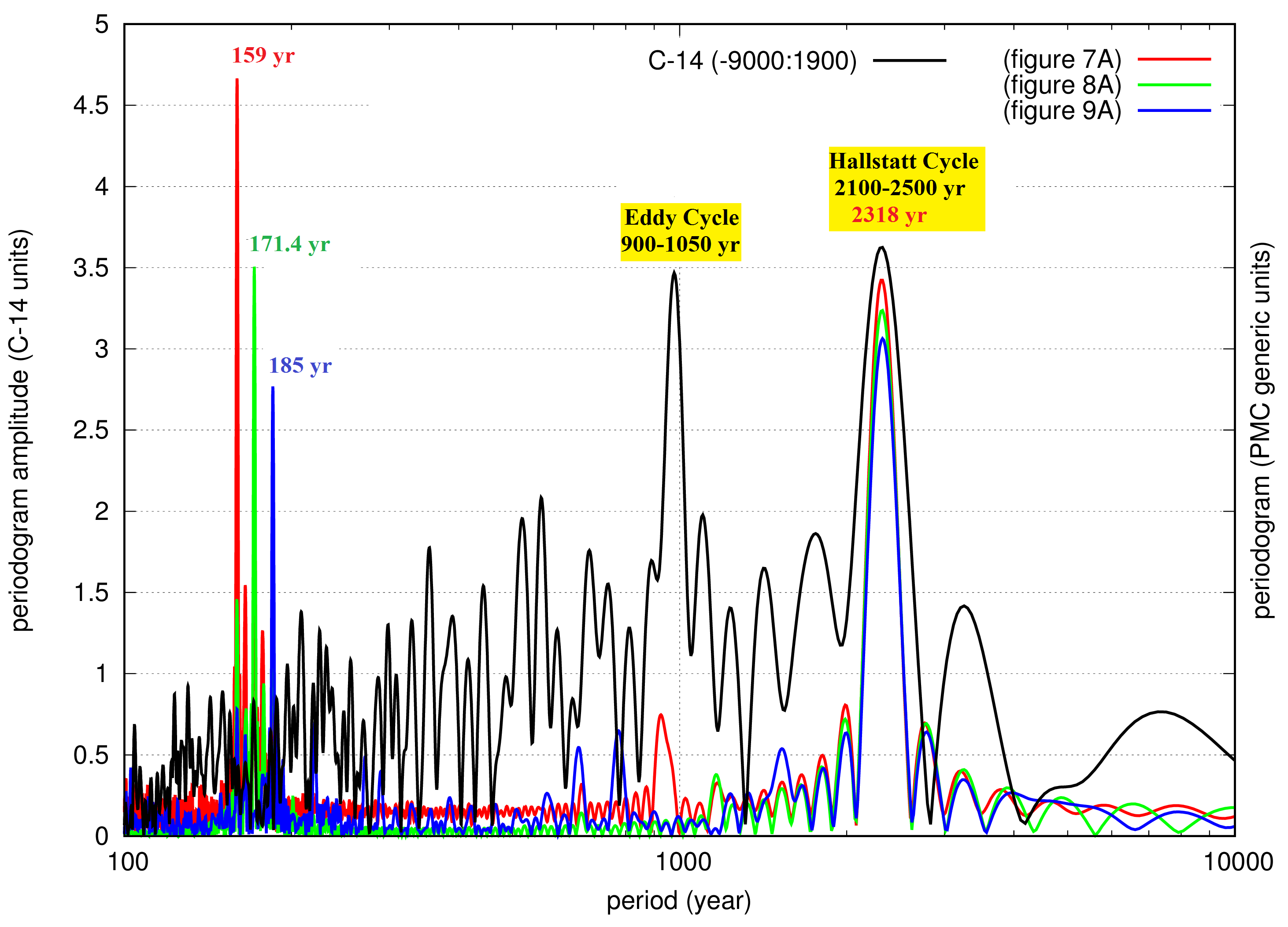}\caption{(Black curve) Periodogram of the $\varDelta^{14}C$ record from -9000
B. C. to 1900 A. D. that is depicted in Figure 1B. (Colored curves)
Same for the Sun-PMC orbital records depicted in Figures 7A, 8A and
9A spanning from 10,000 B. C. to 10,000 A. C.. Note the common spectral
peaks at 2100-2500 yr period which are centered at the orbital resonance
period of $2318$ yr. The three peaks on the left are at about 159
years, 171.4 years and 185 years: the orbital resonances discussed
in Section 3. The Eddy and Hallstatt spectral peaks have a 95\% statistical
confidence with respect to a red-noise background using the Multi
Taper Method, MTM \citep{Ghil}. }
\end{figure}

\begin{figure}[!t]
\centering{}\includegraphics[width=1\textwidth]{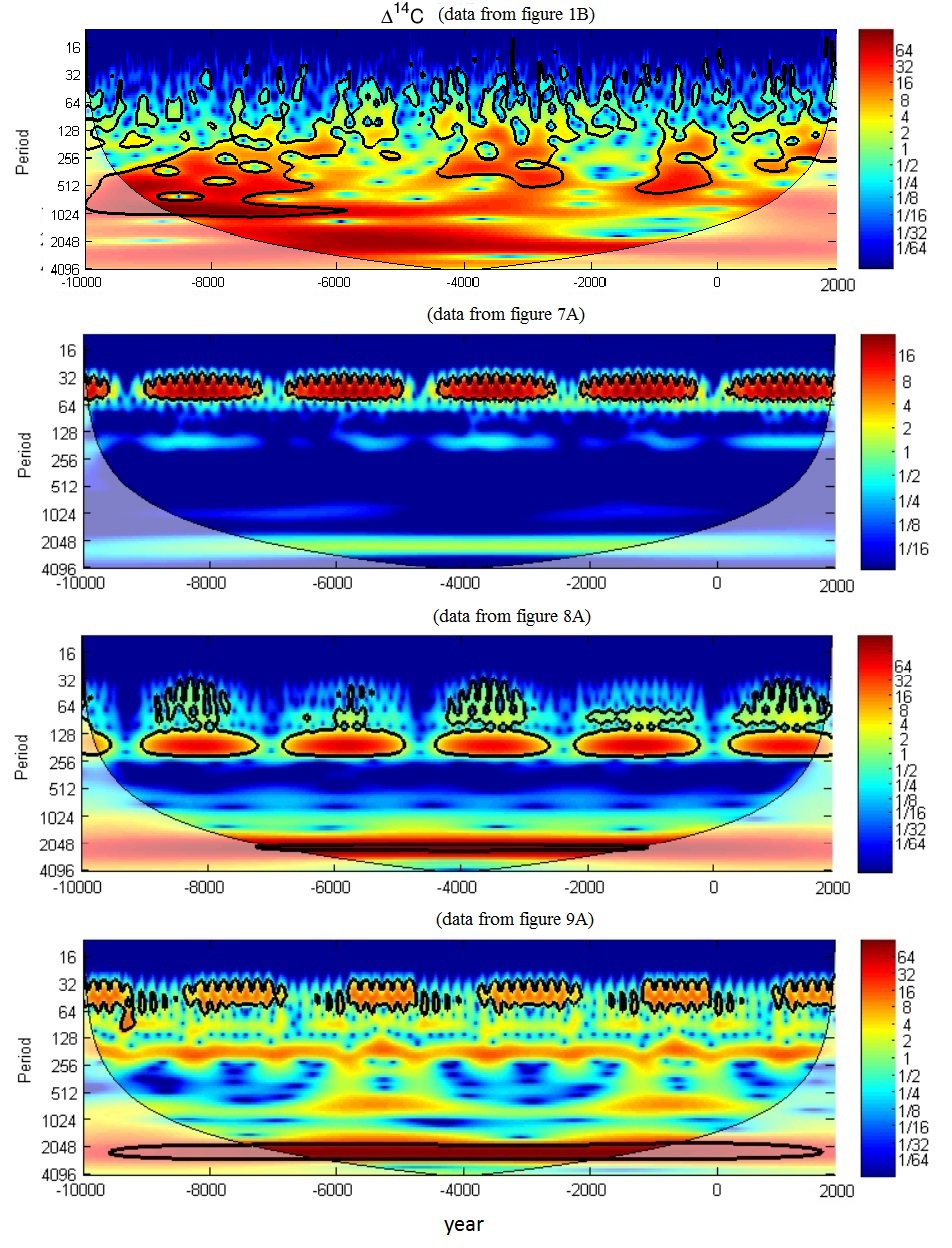}\caption{Continuous wavelet transforms of the records depicted in Figures 1B,
7A, 8A and 9A, respectively.}
\end{figure}

The time sequences depicted in Figures 7-9 clearly put in evidence
a strong oscillation of about 2318 years. Figure 7 reveals the presence
of a major beat frequency with such a period, while Figures 8 and
9 reveal a direct 2318 yr oscillation. Moreover, the phase coincidence
observed in Figure 9 between the Hallstatt oscillation found in $\varDelta^{14}C$
(\textperthousand ) record and in that observed in the astronomical
record suggests that on the 2100-2500 yr time scale the cosmic ray
flux reaching the Earth is higher, when during intervals of about
172 years, within the pericycle-apocycle orbits, the time required
by the PMC to move from the minimum to the maximum distance from the
Sun varies from about 8 to 16 years while the time required by the
PMC to move from the maximum to the minimum distance from the Sun
varies from about 7 to 14 years; on the contrary, the minima of the
radionucleotide production occurred, when the time required by the
PMC to move from the minimum to the maximum distance from the Sun
varies from about 7 to 14 years while the time required by the PMC
to move from the maximum to the minimum distance from the Sun varies
from about 8 to 16 years. 

The power spectra functions depicted in Figure 10 show that the $\varDelta^{14}C$
record depicted in Figure 1B and the Sun-PMC orbital records depicted
in Figures 7-9 share a very large common frequency peak at 2100-2500
year period centered at the stable orbital resonance of 2318 yr. These
spectral peaks have a 95\% statistical confidence against red-noise
background \citep{Ghil}.

Figure 10 shows also that the radionucleotide record presents a significant
900-1050 year Eddy oscillation that has been extensively found in
$^{14}C$, $^{10}Be$ and climate records throughout the Holocene
\citep{Bond,Kerr} and has been modeled involving the orbits of Jupiter
and Saturn and the 11-year solar cycle \citep{Scafetta2012a,Scafetta2014}.
The additional multi-secular minor spectral peaks present in the $\varDelta^{14}C$
record are not further discussed here, but they have been also found
among the planetary harmonics such as the following periods: 104,
130, 150, 171, 185, 208, 354, 500-580 yr \citep[e.g.: ][]{Abreu,Scafetta2014}.
Figure 10 also reveals that the chosen orbital measures present spectral
peaks at about 159 years (from Figure 7A), 171-172 years (from Figure
8A) and 185 years (from Figure 9A), which are also stable orbital
resonances as discussed in Section 3. The 171-172 and 185 yr periods
are visible in the $\varDelta^{14}C$ record although very small,
but they appear well in other solar records \citep[cf: ][]{McCracken2014,Sharp}. 

\begin{figure}[!t]
\centering{}\includegraphics[width=1\textwidth]{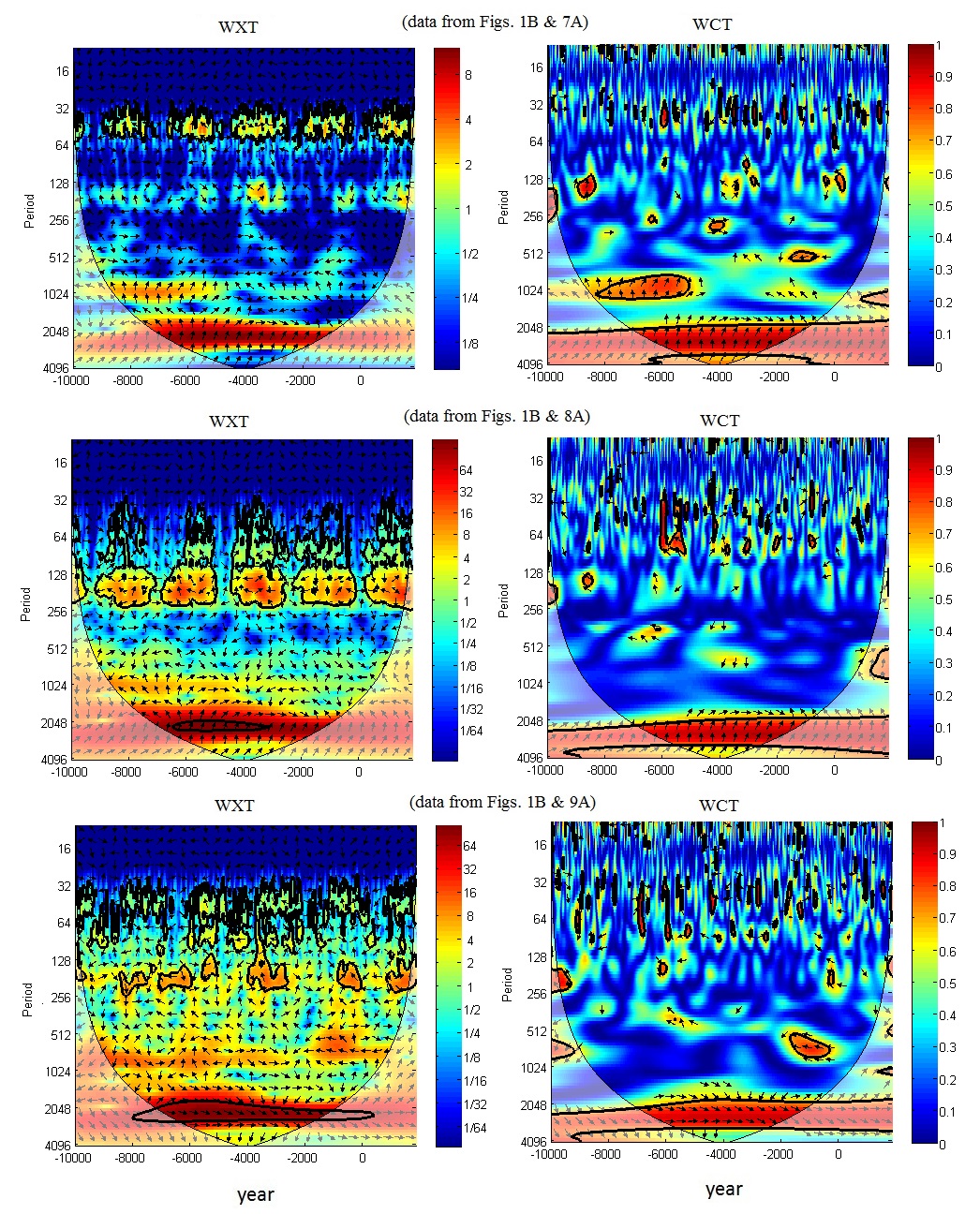}\caption{(Left panels) The cross wavelet transform (XWT) and (right panels)
the wavelet coherence (WTC) between the $\varDelta^{14}C$ record
depicted in Figure 1B and each of the Sun-PMC motion records depicted
in Figures 7A, 8A and 9A, respectively. The red areas surrounded by
the black line satisfy the 95\% confidence level. }
\end{figure}

Figure 11 shows the continuous wavelet transforms of the records depicted
in Figures 1B, 7A, 8A and 9A, respectively. Also these four panels
show that the four records share a significant harmonic at about 2100-2500
yr period. 

Finally, we study the spectral coherence between the $\varDelta^{14}C$
record and the chosen astronomical records. Figure 12 shows in the
left panels the cross wavelet transform (XWT) and in the right panels
the wavelet coherence (WTC) between the $\varDelta^{14}C$ record
depicted in Figure 1B and each of the Sun-PMC motion records depicted
in Figures 7A, 8A and 9A, respectively \citep{Grinsted}. The cross
wavelet transform finds regions in time frequency space where the
time series show high common power. The wavelet coherence finds regions
in time frequency space where the two time series co-vary but does
not necessarily have high power. 

As clearly shown in Figure 12, the six panels demonstrate that the
$\varDelta^{14}C$ record and the chosen astronomical records share
a coherent frequency at about 2100-2500 year period with a 95\% statistical
confidence against red noise background. 

\section{Discussion and Conclusion}

Several experimental evidences demonstrate that throughout the Holocene
the $^{14}C$ atmosphere concentration has varied cyclically in time
\citep[e.g. ][and many others]{Damon1986,Houtermans,Kromer}. An observed
large oscillation has a period of about 2100-2500 years. This oscillation
is known in the scientific literature as the Hallstatt cycle. As discussed
in the Introduction, the presence of a fundamental harmonic at such
a period has been confirmed in numerous studies and found also in
$^{10}Be$ and climate records. For example, recently \citet{McCracken2013}
confirmed an oscillation with period centered between 2300 and 2320
using Fourier amplitude spectrum for GRIP $^{10}Be$, the modelled
estimate of the $^{14}C$ production rate and the modulation function
(in MeV) computed from the EDML and GRIP $^{10}Be$ data, and the
INTCAL09 $^{14}C$ record.

A fundamental scientific issue is to understand the origin of such
an oscillation. It is legitimate to claim that it is an internal climate
or solar oscillation, but in absence of an explicit physical mechanism
this interpretation remains an unproven hypothesis. This leaves open
the possibility for an external astronomical origin of the observed
oscillation. It is observed that the only well-known harmonic generator
of the solar system is provided by the gravitational and electromagnetic
oscillations induced by the revolution of the planets around the Sun.

Thus, we have hypothesized that the Hallstatt oscillation found in
radionucleotide and climatic records could be the result of a specific
orbital resonance within the solar system. A search of the stable
resonances involving the four outer giant planets - Jupiter, Saturn,,
Uranus and Neptune - has determined that, indeed, there exists a major
stable resonance with a period of 2318 years. This stable resonance
is also the only one for period larger than 200 years among those
listed in Table 3. Since this resonance is perfectly coherent to the
Hallstatt oscillation found in radionucleotide and climate records,
this is unlikely a coincidence: we can name this resonance as the
Hallstatt H-resonance of the solar system. 

We have also theoretically determined a large number of additional
stable orbital resonances of the solar system and many of their periods
(e.g. about 20 yr, 44-46 yr, 57-62 yr, 82-97 yr, 159-171-185 yr) are
also typically found in solar, aurora and climate records throughout
the Holocene \citep[e.g.: ][and many others]{Ogurtsov,McCracken2014,Sharp,ScafettaWillson2013a,Scafetta2014,Vaquero}.

Inspired by the \citet{Milankovitch-2}'s theory linking the variation
of the Earth's orbit eccentricity to the glacial cycles, we test whether
the Hallstatt cycle could derive from, and therefore be revealed by,
the overall variation of the circularity of the solar system disk
that could eventually modulate the solar wind intensity and direction
and therefore also the incoming cosmic ray flux and the interplanetary
dust concentration around the Earth. We chose to study the orbit of
the planetary mass center (PMC) relative to the Sun and used the instantaneous
eccentricity vector function \citep[e.g.][]{Mungan} applied to the
Sun-PMC orbit to determine the eccentricity variation of this virtual
planet from 13,000 B. C. to 17,000 A. D. Using spectral analysis we
have demonstrated that this observable presents a significant oscillation
with a 2318 yr period together with a number of already known oscillations
associated to the orbital periods of the planets at scale shorter
than 200 years. Figure 4 stresses that the 2318 yr period peak is
the most relevant in the spectral range between 200 and 10,000 yr
indicating that this oscillation dominates this time scale range.
Thus, there exists a rhythmic contraction and expansion pattern of
the solar system induced by the planets; this pulse is spectrally
coherent to the Hallstatt oscillation found in nucleotides and climate
records. 

\begin{figure}[!t]
\centering{}\includegraphics[width=1\textwidth]{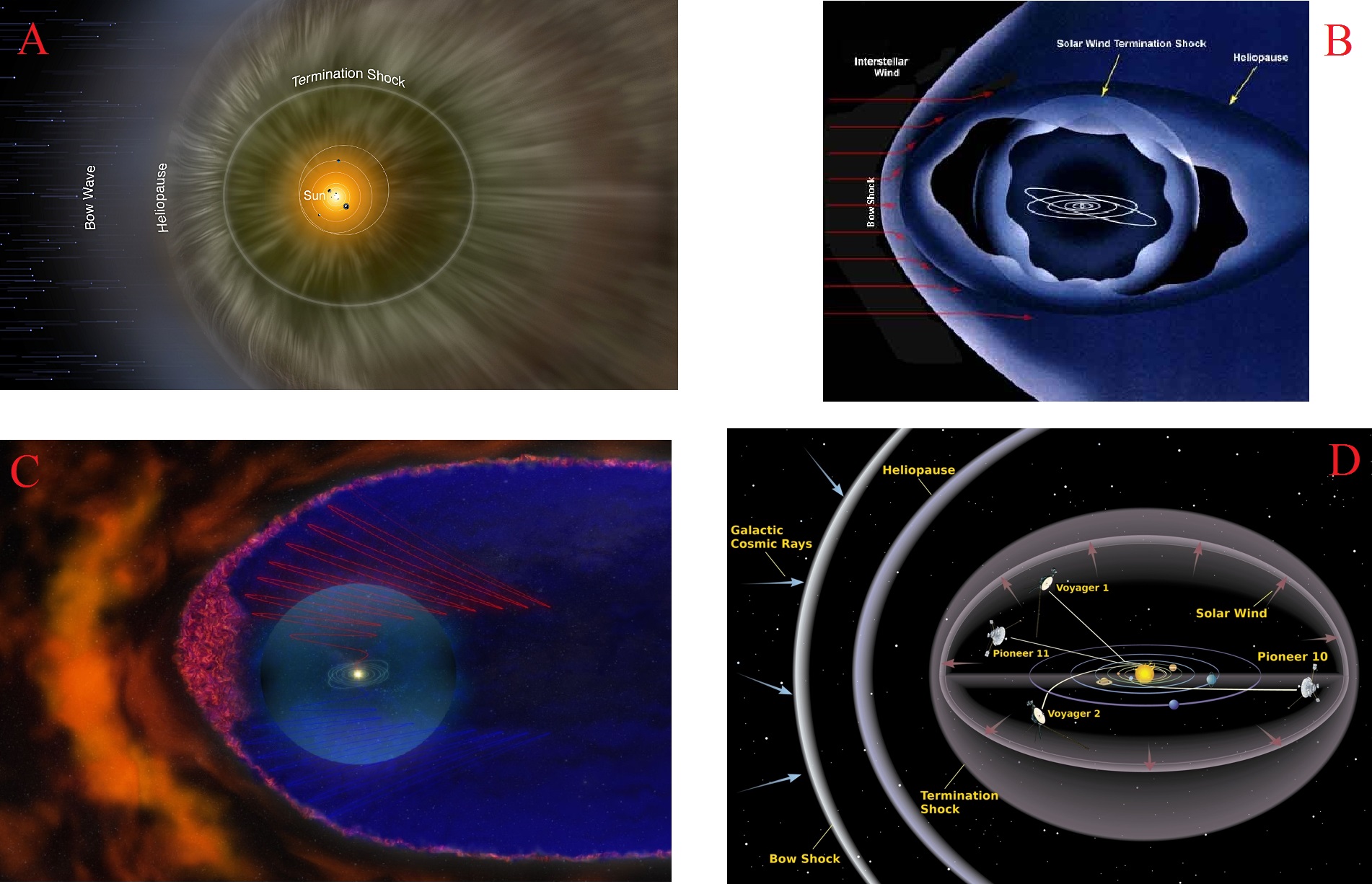}\caption{Artist representations of the heliosphere of the solar system with
highlighted the solar wind termination shock, the heliopause, the
bow shock and the incoming interstellar wind, that is the cosmic ray
flux which is mostly deflected at the heliopause. These illustrations
shows how the sun's activity pushes out cosmic radiation from outside
of the solar system. These and other artist representations of the
heliosphere have been published by NASA \citep[credit to ][and others]{Howell,Phillips}. }
\end{figure}

\begin{figure}[!t]
\centering{}\includegraphics[width=1\textwidth]{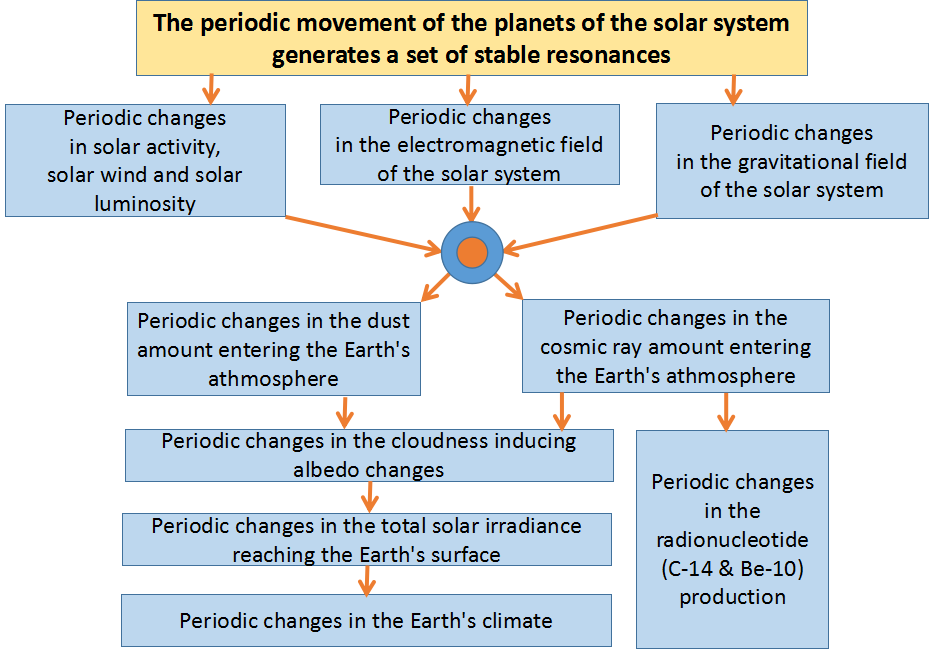}\caption{Schematic flow chart explaining a chain of mechanisms linking the
periodic movement of the planets around the Sun with periodic changes
in solar activity and climate change. }
\end{figure}

In particular, we found a $\pi/2$ phase shift between the 2100-2500
yr curves present in the variation of $^{14}C$ record and the solar
system eccentricity function. Thus, on the Hallstatt-cycle time scale
a larger production of radionucleotide particles, i.e. the occurrence
of a stronger cosmic rays flux toward the inner region of the solar
system, occurs while the Sun-PMC orbit evolves from statistically
more elliptical shapes ($e\approx0.598$) to statistically more circular
ones ($e\approx0.590$): that is while the system is slowly imploding
or bursting inward. Analogously, a smaller production of radionucleotide
particles, i.e. the occurrence of a weaker cosmic ray flux toward
the inner region of the solar system, occurs while the Sun-PMC orbit
evolves from statistically more circular shapes ($e\approx0.590$)
to statistically more elliptical ones ($e\approx0.598$): that is
while the system is slowly exploding or bursting outward.

Finally, to better identify an astronomical proxy able to greatly
stress the 2318-yr H-resonance, we analyzed how the pericycles and
apocycles of the Sun-PMC orbits evolve \citep{Piovan}. We found that
the time series of the periods of these orbits are characterized by
a very prominent 2318 yr oscillation that is perfectly coherent with
the Hallstatt oscillation found in the investigated $\varDelta^{14}C$
record with a statistical confidence above 95\%. These orbital proxies
are also characterized by prominent 159, 171-172, and 185 yr oscillations,
which correspond to other stable resonances of the solar system. Major
harmonics within this spectral range, which was first identified by
\citet{Jose}, is found in long solar activity proxy records \citep{Solanki,Steinhilber}
and in long historical aurora records \citep{ScafettaWillson2013a}.
The coherence at this time scale between our model and the data is
revealed in Figure 12 by the WXT methodology. Other periodicities
found in the eccentricity vector function of the Sun-PMC orbit, such
as about the 20, 30, 45, 60, 87 yr periods, are typically found among
solar \citep[cf. ][]{Ogurtsov}, aurora and climate indexes \citep[cf: ][]{Scafetta2010,Scafetta2012c,Scafetta2013,ScafettaWillson2013a}.

We found (e.g. Figure 9) that at the Hallstatt cycle maxima of the
radionucleotide production occurred when the time required by the
PMC to move from the minimum to the maximum distance from the Sun
varies from about 8 to 16 years, while the time required to move from
the maximum to the minimum distance varies from about 7 to 14 years
throughout the full pericycle-apocycle pattern. On the contrary, the
minima of the radionucleotide production occurred when the time required
by the PMC to move from the minimum to maximum distance from the Sun
varies from about 7 to 14 years while the time required by the PMC
to move from the maximum to the minimum distance from the Sun varies
from about 8 to 16 years. This suggests that, at this time scale,
the cosmic ray flux increases (decreases) during periods of slower
(faster) expansion and faster (slower) contraction of the Sun-PMC
orbit wobbling. Thus, a faster expansion of the solar system prevents
cosmic rays to enter in its inner regions, while a faster contraction
favors a larger incoming of cosmic rays. This suggests a rhythmic
modulation of the geometry of the heliopausa and/or of solar wind
termination shock and, therefore, of the heliospheric magnetic field
\citep{Owen}, which in the former case should become larger while
in the latter should become smaller, inducing a larger or smaller
deviation of the incoming cosmic rays, respectively.

Figure 13 shows a graphical representation of the heliosphere of the
solar system with highlighted the solar wind termination shock, the
heliopause, the bow shock and the incoming interstellar wind. A reader
should note that the heliosphere is not stationary relative to the
incoming cosmic rays, but it is wobbling like the Sun-PMC orbits that
we have studied in this work. Since the planetary system is within
the heliosphere this means that the interaction of the planets with
the inner heliosphere can be relevant. In particular, magnetic field
reconnections can capture and redirect solar wind shaping the heliosphere,
which modulates the cosmic ray flux. The imploding-exploding dynamics
revealed in our record could easily modulate the solar wind termination
shock surface and, therefore, modulate the incoming cosmic ray flux.
\citet{Potgieter} summarizes mechanisms of solar modulation of cosmic
rays in the heliosphere.

A related but complementary mechanism is that the displacements of
the planets could directly or indirectly modulate the amount of interplanetary/cosmic
dust falling on Earth. The cosmic-planetary dust is concentrated within
the disk of the solar system where the Sun-PMC orbit evolves, including
the region surrounding the Earth-Moon system, and regulates the intensity
of the zodiacal light \citep[cf.: ][]{Ermakov,Ermakovb,Ollila}. In
fact, every day from 400 to 10,000 tons of dust enters in the Earth\textquoteright s
atmosphere. The sizes of these particles vary from 0.001 \textmu m
to several hundreds of micrometres and they are mostly made of common
elements such as Fe, Mg, S, Al, Ca, and Na. Because these particles
are very likely also charged by solar wind, they should be subject
not only to gravitational forces but also to magnetic fields and to
the solar wind itself \citep{Divari}. Therefore, interplanetary/cosmic
dust can be driven by the planets \citep{Divari}. Once that these
charged particles enter the Earth's atmosphere, they behave as efficient
condensation nuclei of the atmospheric water vapor, which is a polarized
molecule, in particular Mg, S, and Na, and also help in forming clouds
\citep[cf.: ][]{Ermakov,Ermakovb,Ollila}. The Sun-PMC wobbling could
more easily disperse this dust away from the inner region of the solar
system when its orbit expands fast and contracts slower as depicted
in Figure 9 inducing the formation of less clouds on the Earth, and
vice versa. Indeed, while long records of interplanetary/cosmic dust
falling on Earth are not available to test this hypothesis, a record
of historically recorded meteorite falls in China from 619 to 1943
A. D. has revealed harmonics such as 10.5, 14-15, 30 and 60-63 year
oscillations \citep{Scafetta2012c,Yu}, which are also found among
the main harmonics of the eccentricity function of the Sun-PMC orbit
and among the stable resonances of the solar system (see Table 1).
A modulation of the interplanetary/cosmic dust density surrounding
the Earth-Moon system driven by the solar wind and the planets' magnetospheres
may also contribute to explain why the Hallstatt cycle, and a number
of other oscillations revealed in Figure 4, are also observed in climate
records \citep[e.g.: ][]{Levina1993,OBrien1995}. Further research
will better clarify the specific physical mechanisms involved in these
processes. 

In conclusion, our results clearly suggest that the velocity of the
displacements of the Jovian planets (Jupiter, Saturn, Uranus and Neptune),
which mainly determine the inertial motion of the Sun on the long
time scales, also influences the solar activity and the intensity
of its solar wind, and/or modifies the structure of the heliosphere.
The latter would then modulate the incoming cosmic ray flux that produces
radiocarbon and/or interplanetary/cosmic dust concentration and, simultaneously,
they would regulate the Earth's climate by modulating the cloud system:
see the schematic flow chart depicted in Figure 14. A possible chain
of the involved mechanisms has been suggested by some authors \citep{Kirby,Ollila,Scafetta2012c,Scafetta2013,Shaviv,Svensmark2,Svensmark,Tinsley}.
Moreover, the fact that a specific set of planetary resonances may
be modulating eliospheric, solar and climatic records may suggest
why these records appear linked at multiple time scales, and even
at the short ones \citep[e.g.: ][]{Scafetta2004,ScafettaWest2006,Scafetta2009}.

Although it is still uncertain how the planets could influence solar
activity and/or the cosmic ray flux and/or the dust reaching the Earth,
a planetary origin of solar and climate oscillations, which has been
proposed since antiquity, has recently received a renewed attention
in the scientific literature. Planetary theories of climate variations
were widespread in ancient times and, in more recent times, a planetary
theory of solar variation was proposed by \citet{Wolf1859} to explain
the 11-year solar cycle. Wolf hypothesized that the just discovered
11-year solar cycle could emerge from a combined influence of Venus,
Earth, Jupiter and Saturn, which has been recently confirmed \citep{Hung,Scafetta2012a,Scafetta(2012b),Wilson}.
\citet{Stefani} speculated that the tidal oscillation of 11.07 years
induced by the Venus\textendash Earth\textendash Jupiter system may
lead to a 1:1 resonant excitation of the oscillation of the \textgreek{a}-effect.
In general, a planetary origin of solar and climate oscillations is
based on numerous empirical evidences at multiple time scales and
some preliminary physical explanations \citep[e.g.: ][]{Abreu,Charv=0000E1tov=0000E1,Cionco,Hung,Jakubcov=0000E1,Jose,McCracken2013,McCracken2014,M=0000F6rner2013,M=0000F6rner2015,Puetz,Salvador,Scafetta2010,Scafetta2012a,Scafetta(2012b),Scafetta2013,Scafetta2014,Scafetta2016,ScafettaWillson2013a,ScafettaandWillson(2013b),Sharp,Solheim,TanCheng,Tattersall,Wilson}. 

Since the 19th century, the planetary theory of solar and climate
variations has also received a number of critiques \citep[some of the most recent critical studies include: ][]{Callebaut,Cameron,Cauquoin-1,Holm2014a,Holm2015,Poluianov2014,Smythe}.
However, several rebuttals of the critiques have also been published.
The rebuttals can be summarized as follows: i) the Sun can react to
a planetary tidal forcing because it is a nuclear fusion generator
that might greatly amplify the modest gravitational tidal effect \citep[e.g.: ][]{Scafetta(2012b),Wolff};
ii) an additional electromagnetic coupling could link the Sun to the
planets throughout the solar wind \citep[e.g.: ][]{ScafettaandWillson(2013b)};
iii) the solar-climate physics occurs throughout some heliosphere
dynamics \citep[e.g.:  ][]{ScafettaandWillson(2013b),ScafettaWillson2013a};
iv) the coupling between some astronomical and the solar-climate harmonics
is very good when the appropriate astronomical proxies that takes
into account multiple planets are constructed \citep[e.g.: ][]{Scafetta2014,Scafetta2016,Sharp,Wilson};
v) the spectral coherence at the given harmonics is statistically
significant above 95\% when the calculations are done correctly and
once the limits of the used analysis algorithms are properly considered
\citep[see also:][]{Scafetta2014,Scafetta2016}; Monte Carlo techniques
used to test the likelihood that multiple frequencies in solar records
match planetary records show unambiguously that this probability is
lower than $10^{-4}$ \citep[e.g.: ][]{Abreu2014,ScafettaandWillson(2013b)}. 

We also noted that once challenged, some critical authors responded
by contradicting their previous claims. For example, \citet{Callebaut}
argued against a planetary origin of the solar oscillation by also
claiming that the five major identified solar periodicities - Schwabe
($\sim$11 yr), Hale ($\sim$22 yr), Gleissberg ($\sim$88 yr), Suess
($\sim$203-208 yr) and Hallstatt ($\sim$2300-2400 yr) cycles - were
never successfully reproduced by papers advocating planetary influences
on solar variability. However, when their claim was challenged \citep[e.g.: ][]{ScafettaHumlum2013},
\citet{Callebaut2013} acknowledged that \textit{``it is well-known
that there are some periodicities that are common to solar activity
and planetary motions''} which make his further critique very weak
as explained in \citet{ScafettaHumlum2013}. Similarly, \citet{Holm2014a}
argued that using windowed periodograms no spectral coherence between
temperature records and the speed of the solar center of mass could
be found at given frequencies such as at 20 yr and 60 yr periods.
However, after that \citet{Scafetta2014} demonstrated that Holm used
improperly the windowed periodogram, that is he used it with too short
window segments to detect the signals, \citet{Holm2015} acknowledged
that \textit{``it is not hard to produce high coherence estimates
at periods around 15\textendash 22 and 50\textendash 60 years between
these data sets.''} Rebuttals of \citet{Holm2015}'s further critiques
were presented in \citet{Scafetta2016}.

Although several issues remain open to further investigations, the
published scientific literature provides several evidences that solar
and climate records are characterized by periodicities that are common
to planetary motions at multiple time scales from a few months to
several millennia. Moreover, no alternative explanations of these
oscillations have been proposed by the critics. In other words, an
alternative theory explaining the observed oscillations simply does
not exist. The methodologies and results of the present paper contribute
to this discussion showing compelling evidences that also the long
Hallstatt (2100-2500 yr) oscillation likely has an astronomical origin
linked to the internal dynamics of the solar system and its stable
resonances such as those produced by Jupiter, Saturn, Uranus and Neptune
at periods of 159 yr, 171 yr, 185 yr and 2318 yr. From Figure 5, the
next Hallstatt minimum in the $^{14}C$ cosmogenic radioisotopes will
occur around 2804 A. D. and the next maximum around 3963 A. D..

\subsection*{Acknowledgment}

The authors would like to thank the two referees for very useful suggestions
and comments.

\newpage{}

\newpage{}

\begin{thebibliography}{Cameron, R. H., Schüssler(2013)}
\bibitem[Abreu et al.(2012)]{Abreu}Abreu, J. A., Beer, J., Ferriz-Mas,
A., McCracken, K. G., Steinhilber, F.: Is there a planetary influence
on solar activity? Astron. Astrophys., 548, A88, 2012.

\bibitem[Abreu et al.(2014)]{Abreu2014}Abreu, J. A., Albert, C.,
Beer J., Ferriz-Mas A., McCracken K. G., Steinhilbe F.: Response to:
\textquotedblleft Critical Analysis of a Hypothesis of the Planetary
Tidal Influence on Solar Activity\textquotedblright . Solar Physics
289, 2343\textendash 2344, 2014.

\bibitem[Adolphi et al.(2014)]{Adolphi}Adolphi, F., Muscheler, R.,
Svensson, A., Aldahan, A., Possnert, G., Beer, J., Sjolte, J., Bj{\"o}rck,
S., Matthes, K., Thi{\'e}blemont, R.: Persistent link between solar
activity and Greenland climate during the Last Glacial Maximum. Nature
Geoscience 7, 662\textendash 666, 2014.

\bibitem[Bard et al.(1997)]{Bard1997}Bard, E., Raisbeck, G. M., Yiou,
F., Jouzel, J.: Solar modulation of cosmogenic nuclide production
over the last millennium: comparison between $^{14}C$ and $^{10}Be$
records. Earth and Planetary Science Letters, 150(3\textendash 4),
453\textendash 462, 1997.

\bibitem[Bard et al.(2000)]{Bard}Bard, E., Raisbeck, G., Yiou, F.,
Jouzel, J.: Solar irradiance during the last 1200 years based on cosmogenic
nuclides. Tellus 52B, 985\textendash 992, 2000.

\bibitem[Bray(1968)]{Bray}Bray, J. R.: Glaciation and solar activity
since the fifth century B.C and the solar cycle. Nature, 220, 672-674,
1968.

\bibitem[Bond et al.(2001)]{Bond}Bond, G., Kromer, B., Beer, J.,
Muscheler, R., Evans, M.N., Showers, W., Hoffmann, S., Lotti-Bond,
R., Hajdas, I., Bonani, G.: Persistent solar influence on North Atlantic
climate during the holocene. Science 294, 2130\textendash 2136, 2001.

\bibitem[Cameron, R. H., Schüssler(2013)]{Cameron}Cameron, R. H.,
Schüssler, M.: No evidence for planetary influence on solar activity.
Astron. \& Astrophys., 557, A83, 2013.

\bibitem[Cauquoin et al.(2014)]{Cauquoin-1}Cauquoin, A., Raisbeck,
G. M., Jouzel, J., Bard, E. (ASTER Team): No evidence for planetary
influence on solar activity 330 000 years ago. Astron. \& Astrophys.,
561, A132, 2014.

\bibitem[Callebaut et al.(2012)]{Callebaut}Callebaut, D. K., de Jager,
C., Duhau, S.: The influence of planetary attractions on the solar
tachocline. Journal of Atmospheric and Solar\textendash Terrestrial
Physics 80, 73\textendash 78, 2012.

\bibitem[Callebaut et al.(2013)]{Callebaut2013}Callebaut, D. K.,
de Jager, C., Duhau, S.: Reply to \textquotedblleft The influence
of planetary attractions on the solar tachocline\textquotedblright{}
by N. Scafetta, O. Humlum, J.E. Solheim, K. Stordahl. Journal of Atmospheric
and Solar\textendash Terrestrial Physics 102, 372, 2013.

\bibitem[Charv{\'a}tov{\'a}(2000)]{Charvatova}Charv{\'a}tov{\'a},
I.: Can origin of the 2400-year cycle of solar activity be caused
by solar inertial motion? Ann. Geophysicae, 18, 399\textendash{} 405,
2000.

\bibitem[Charv{\'a}tov{\'a}(2009)]{Charv=0000E1tov=0000E1}Charv{\'a}tov{\'a},
I.: Long-term predictive assessments of solar and geomagnetic activities
made on the basis of the close similarity between the solar inertial
motions in the intervals 1840\textendash 1905 and 1980\textendash{}
2045. New Astron., 14, 25-30, 2009.

\bibitem[Cionco and Soon(2014)]{Cionco} Cionco, R. G., Soon, W.:
A phenomenological study of the timing of solar activity minima of
the last millennium through a physical modeling of the Sun-Planets.
New Astronomy, 34, 164-171, 2015.

\bibitem[Creer(1988)]{Creer}Creer, K. M.: Geomagnetic field and radiocarbon
activity through Holocene time. In: Secular, Solar and Geomagnetic
Variations in the last 10000 years. (Eds) Stephenson, F. R. andWolfendale,
A. W., Kluwer, Dordrecht, p. 381\textendash 397, 1988.

\bibitem[Czymzik et al.(2016)]{Czymzik}Czymzik, M., Muscheler, R.,
Brauer, A.: Solar modulation of flood frequency in central Europe
during spring and summer on interannual to multi-centennial timescales.
Climate of the Past, 12(3), 799-805, 2016.

\bibitem[Damon(1988)]{Damon1988}Damon, P. E.: Production and decay
radiocarbon and its modulation by geomagnetic field-solar activity
changes with possible implications for global environment. In: Secular,
Solar and Geomagnetic Variations in the Last 10000 years, (Eds) Stephenson,
F. R. and Wolfendale, A. W., Kluwer, Dordrecht, p. 267\textendash 285,
1988.

\bibitem[Damon and Linick(1986)]{Damon1986}Damon, P. E., Linick,
T. W.: Geomagnetic-heliomagnetic modulation of atmospheric radiocarbon
production. Radiocarbon, 28(2A), 266\textendash 278, 1986.

\bibitem[Damon et al.(1990)]{Damon1990}Damon, P. E., Cheng, S., Linick,
T. W.: Fine and hyperfine structure in the spectrum of secular variations
of atmospheric $^{14}C$. Radiocarbon, 31(3), 704\textendash 718,
1990.

\bibitem[Damon and Sonett(1992)]{Damon1992}Damon, P. E., Sonett,
C. P.: Solar and terrestrial components of the atmospheric 14C variation
spectrum. In: The Sun in Time, (Eds) Sonett, C. P., Giampapa, M. S.,
Mathews, M. S., The Univesity of Arizona Press, Tucson, p. 360\textendash 388,
1992.

\bibitem[Damon and Jirikowi{\'c}(1992)]{Damon1992b}Damon, P. E.,
Jirikowi{\'c}, J. L.: Radiocarbon evidence for low frequency solar
oscillation. In: Rare Nuclear Processes, (Ed) Povinec, P., Proc. 14th
Europhysics Conf. on Nuclear Physics, Word Scientific Publishing Co,
Singapore, p. 177\textendash 202, 1992.

\bibitem[Dansgaard et al.(1984)]{Dansgaard}Dansgaard, W., Johnsen,
S. J., Clausen, H. B., Dahl-Jensen, D., Gunderstrup, N., Hammer, C.,
Oeschger, H.: North Atlantic climate oscillations revealed by deep
Greenland ice core. In: Climate Processes and Climate Sensivity, (Eds)
Hansen, J. E. and Takahashi, T., AGU, Washington, D. C., p. 288\textendash 298,
1984.

\bibitem[Darwin(1902)]{Darwin}Darwin, G.: Tides. (Encyclopædia Britannica,
Ninth Edition, 1875\textendash 89) 1902. http://www.1902encyclopedia.com/T/TID/tides.html

\bibitem[Davis and Bohling(2001)]{Davis}Davis, J. C., Bohling, G.:
The search for patterns in ice-core temperature curves. In: Gerhard,
L.C., Harrison, W.E., Hanson, B.M. (Eds.), Geological Perspectives
of Global Climate Change, pp. 213\textendash 229, 2001.

\bibitem[Dergachev and Chistyakov(1995)]{Dergachev}Dergachev, V.,
Chistyakov, V.: Cosmogenic radiocarbon and cyclical natural processes.
Radiocarbon, 37(2), 417\textendash 424, 1995.

\bibitem[Divari(1966)]{Divari}Divari, N. B.: Charged Dust Particles
in Interplanetary Space. Soviet Astronomy, 10, 151-154, 1966.

\bibitem[Doodson(1921)]{Doodson}Doodson, A. T.: The Harmonic Development
of the Tide-Generating Potential. Proceedings of the Royal Society
of London. Series A, 100 (704) , 305-329, 1921.

\bibitem[Elsasser et al.(1956)]{Elsasser}Elsasser, W., Ney, E. P.,
Winckler, J. R.: Cosmic ray intensity and geomagnetism. Nature, 178,
1226\textendash 1227, 1956.

\bibitem[Ermakov et al.(2009a)]{Ermakov}Ermakov, V. I., Okhlopkov,
V. P., Stozhkov, Yu. I.: The impact of cosmic dust on the earth\textquoteright s
climate. Moscow University Physics Bulletin, 64(2), 214\textendash 217,
2009a.

\bibitem[Ermakov et al.(2009b)]{Ermakovb}Ermakov, V. I., Okhlopkov,
V. P., Stozhkov, I. Yu.: Influence of cosmic rays and cosmic dust
on the atmosphere and Earth\textquoteright s climate. Bulletin of
Russian Academy of Sciences: Physics, 434-436, 2009b.

\bibitem[Fairbridge(1984)]{Fairbridge}Fairbridge, R. W., 1984. Planetary
periodicities and terrestrial climate stress. In M{\"o}rner and Karlén
(1984), 509-520, 1984.

\bibitem[Fairbridge and Sanders(1987)]{Fairbridge2}Fairbridge, R.
W., Sanders, J. E.: 1987. The Sun\textquoteright s orbit, AD 250-2050:
basis for new perspectives on planetary dynamics and Earth-Moon linkage.
In Rampino et al. (1987), 446-471, bibliography 475-541, 1987.

\bibitem[Folkner et al.(2014)]{Folkner}Folkner, W. M., Williams,
J. G., Boggs, D. H., Park, R. S., Kuchynka, P.: The Planetary and
Lunar Ephemerides DE430 and DE431. IPN Progress Report 42-196, 2014.
http://ipnpr.jpl.nasa.gov/progress\_report/42-196/196C.pdf 

\bibitem[Folkner(2014)]{Folkner-1}Folkner, W. M.: Planetary Ephemeris
DE432. Jet Propulsion Laboratory Memorandum IOM 392R-14-003, 2014.

\bibitem[Gervais(2016)]{Gervais}Gervais, F.: Anthropogenic CO2 warming
challenged by 60-year cycle. Earth-Science Reviews, 155, 129-135,
2016.

\bibitem[Ghil et al.(2002)]{Ghil}Ghil, M., Allen, M. R., . Dettinger,
M. D., Ide, K., Kondrashov, D., Mann, M. E., Robertson, A. W., Saunders,
A., Tian, Y., Varadi, F., Yiou, P.: Advanced Spectral Methods for
Climatic Time Series. Reviews of Geophysics 40, 1003, 2002. SSA-MTM
Toolkit for spectral analysis (http://research.atmos.ucla.edu/tcd//ssa/). 

\bibitem[Gregori(2002)]{Gregori2002}Gregori, G. P.: Galaxy \textendash{}
Sun \textendash{} Earth relations. The origin of the magnetic field
and of the endogenous energy of the Earth, with implications for volcanism,
geodynamics and climate control, and related items of concern for
stars, planets, satellites, and other planetary objects. A discussion
in a prologue and two parts. Beiträge zur Geschichte der Geophysik
und Kosmischen Physik, Band 3, Heft 3, 471 pp., 2002.

\bibitem[Grinsted et al.(2004)]{Grinsted}Grinsted, A., Moore, J.
C., Jevrejeva, S.: Application of the cross wavelet transform and
wavelet coherence to geophysical time series, Nonlin. Process. Geophys.,
11, 561566, 2004.

\bibitem[Goslar et al.(1999)]{Goslar}Goslar, T., Wohlfarth, B., Bj{\"o}rck,
S., Possnert, G., Bj{\"o}rck, J.: Variations of atmospheric $^{14}C$
concentrations over the Allerød- Younger Dryas transition. Climate
Dynamics, 15, 29\textendash 42, 1999.

\bibitem[Hood and Jirikowi{\'c}(1990)]{Hood}Hood, L. L., Jirikowi{\'c},
J. L.: Recurring variations of probable solar origin in the atmospheric
$^{14}C$ time record. Geophys. Res. Letts., 17, 85, 1990.

\bibitem[Hoyt and Schatten(1997)]{Hoyt}Hoyt, D. V., Schatten, K.
H.: The Role of the Sun in the Climate Change. Oxford Univ. Press,
New York, 1997.

\bibitem[Holm(2014)]{Holm2014a}Holm, S.: On the alleged coherence
between the global temperature and the Sun\textquoteright s movement.
J. Atmos. Sol.-Terr. Phys., 110-111, 23-27, 2014a.

\bibitem[Holm(2015)]{Holm2015}Holm, S.: Prudence in estimating coherence
between planetary, solar and climate oscillations. Astrophys. Space
Sci., 357:106, 1-8, 2015.

\bibitem[Howell(2014)]{Howell}Howell, E.: 2014. Weak Sun poses radiation
risk for Mars-bound astronauts. Space.com, issued Nov 08, 2014. 

\bibitem[Houtermans(1971)]{Houtermans}Houtermans, J. C.: Geophysical
interpretation of Bristlecone pine radiocarbon measurements using
a method of Fourier analysis of unequally spaced data. Ph.D. Thesis,
Univ. of Bern, 1971.

\bibitem[Humlum et al.(2011)]{Humlum}Humlum, O., Solheim, J.-E.,
Stordahl, K.: Identifying natural contributions to late Holocene climate
change. Global and Planetary Change, 79(1), 145-156, 2011.

\bibitem[Hung(2007)]{Hung}Hung, C.-C.: Apparent relations between
solar activity and solar tides caused by the planets. NASA report/TM-2007-214817
(2007). Available at http://ntrs.nasa.gov/search.jsp?R=20070025111

\bibitem[Kerr(2001)]{Kerr}Kerr, R. A.: A variable sun paces millennial
climate. Science 294, 1431\textendash 1433, 2001.

\bibitem[Kirby(2007)]{Kirby} Kirby, J.: Cosmic Rays and Clouds. Surveys
in Geophysics, 28, 333-373, 2007.

\bibitem[Kromer et al.(1998)]{Kromer}Kromer, B., Spurk, M., Remmele,
S., Barbetti, M., Toniello, V.: Segments of atmospheric $^{14}C$
change as devided from Late Glacial and Early Holocene floating tree-ring
series. Radiocarbon, 40(1), 351\textendash 358, 1998.

\bibitem[Jakubcov{\'a} and Pick(1986)]{Jakubcov=0000E1}Jakubcov{\'a},
I., Pick, M.: The planetary system and solar-terrestrial phenomena.
Stud. Geophys. Geod., 30, 224-235, 1986.

\bibitem[Jose(1965)]{Jose}Jose, P. D.: Sun\textquoteright s motion
and sunspots. Astron. J., 70, 193-200, 1965.

\bibitem[Lal(1988)]{Lal}Lal, D.: Theoretically expected variations
in the terrestrial cosmicray production rates of isotopes. In: Solar-terrestrial
relationships and the Earth environment in the last millennia, (Ed)
Castagnoli, G. C., North-Holland Press, Amsterdam, p. 216\textendash 233,
1988.

\bibitem[Levina and Orlova(1993)]{Levina1993}Levina, T. P., Orlova,
L. A.: Holocene climatic rhythms of southern West Siberia. Russian
Geology and Geophysics, 34, 36\textendash 51, 1993.

\bibitem[Loehle and Scafetta(2011)]{Loehle}Loehle, C., Scafetta,
N.: Climate Change Attribution Using Empirical Decomposition of Climatic
Data. The Open Atmospheric Science Journal 5, 74-86, 2011.

\bibitem[Manzi et al.(2012)]{Manzi}Manzi, V., Gennari R., Lugli S.,
Roveri M., Scafetta N., Schreiber C.: High-frequency cyclicity in
the Mediterranean Messinian evaporites: evidence for solar-lunar climate
forcing. Journal of Sedimentary Research 82, 991-1005, 2012.

\bibitem[Marcott et al.(2013)]{Marcott}Marcott, S. A., Shakun, J.
D., Clark, P. U., Mix, A. C.: A reconstruction of regional and global
temperature for the past 11,300 years. Science, 339(6124), 1198-1201,
2013.

\bibitem[Mayewski et al.(2004)]{Mayewski}Mayewski, P. A., Rohling,
E. E., Stager, J. C., Karlen, W., Maasch, K. A., Meeker, L. D., Meyerson,
E. A., Gasse, F., van Kreveld, S., Holmgren, K., Lee-Thorp, J., Rosqvist,
G. Rack, F., Staubwasser, M., Schneider, R. R. and Steig, E. J.: Holocene
climate variability. Quaternary Research, 62(3), 243-255, 2004.

\bibitem[Mazzarella and Scafetta(2012)]{Mazzarella}Mazzarella, A.,
Scafetta, N.: Evidences for a quasi 60-year North Atlantic Oscillation
since 1700 and its meaning for global climate change. Theoretical
Applied Climatology 107, 599-609, 2012.

\bibitem[McCracken et a.(2001)]{McCracken2001}McCracken, K. G., Dreschhoff,
G. A. M, Smart, D. F. , Shea, M. A.: Solar cosmic ray events for the
period 1561\textendash 1994: 2. The Gleissberg periodicity. J. Geophys.
Res 106.21: 599-21, 2001.

\bibitem[McCracken et al.(2013)]{McCracken2013}McCracken, K. G.,
Beer, J., Steinhilber, F., Abreu, J.: A phenomenological study of
the cosmic ray variations over the past 9400 years, and their implications
regarding solar activity and the solar dynamo. Sol. Phys., 286, 609\textendash 627,
2013. 

\bibitem[McCracken et al.(2014)]{McCracken2014}McCracken, K. G.,
Beer, J., Steinhilber, F., Abreu, J.: Evidence for planetary forcing
of the cosmic ray intensity and solar activity throughout the past
9400 years. Sol. Phys., 286(2), 609\textendash 627, 2014. 

\bibitem[Melchior(1978)]{Melchior}Melchior, P.: The Tides of the
Planet Earth. Pergamon Press, Oxford, 1978. 

\bibitem[Milankovi{\'c}(1930)]{Milankovitch-2}Milankovi{\'c}, M.:
Mathematische Klimalehre und Astronomische Theorie der Klimaschwankungen,
Handbuch der Klimatologie, Band I, Teil A,Berlin, Verlag von Gebrüder
Borntraeger, 1930.

\bibitem[M{\"o}rner(1996)]{morner1996}M{\"o}rner, N.-A.: Global
Change and interaction of Earth rotation, ocean circulation and paleoclimate,
An. Brazilian Acad. Sc., 68 (Supl. 1), 77\textendash 94, 1996.

\bibitem[M{\"o}rner(2013)]{M=0000F6rner2013}M{\"o}rner, N.-A.: Planetary
beat and solar\textendash terrestrial responses. Pattern Recogn. Phys.,
1, 107-116, 2013.

\bibitem[M{\"o}rner(2015)]{M=0000F6rner2015}M{\"o}rner, N.-A.: The
Approaching New Grand Solar Minimum and Little Ice Age Climate Conditions.
Natural Science, 7, 510-518, 2015.

\bibitem[Mortari(2010)]{Mortari}Mortari, R.: I ritmi segreti dell\textquoteright universo,
336 pp., (3rd ed.) Aracne editrice s.r.l., Roma. The 1st ed. appeared
in 1988, the 2nd ed. in 1999, 2010.

\bibitem[Mungan(2005)]{Mungan}Mungan, C. E.: Another comment on \textquoteleft Eccentricity
as a vector\textquoteright . Eur. J. Phys., 26, L7-L9, 2005.

\bibitem[O\textquoteright Brien(1979)]{OBrien}O\textquoteright Brien,
K.: Secular variations in the production of cosmogenic isotopes in
the Earth\textquoteright s atmosphere. J. Geophys. Res., 84, 423\textendash{}
431, 1979.

\bibitem[O\textquoteright Brien et al.(1995)]{OBrien1995}O\textquoteright Brien,
S. R., Mayewski, P. A., Meeker, L. D., Meese, D. A., Twickler, M.
S., Whitlow, S. I.: Complexity of Holocene climate as reconstructed
from a Greenland ice core. Science, 270, 1962\textendash 1964, 1995.

\bibitem[Oeschger et al.(1975)]{Oeschger}Oeschger, H., Siegenthaler,
U., Schotterer, U., Gugelmann, A.: A box diffusion model to study
the carbon dioxide exchange in nature. Tellus, 27, 168\textendash 192,
1975.

\bibitem[Ollila(2015)]{Ollila}Ollila, A.: Cosmic theories and greenhouse
gases as explanations of global warming. Journal of Earth Sciences
and Geotechnical Engineering 5(4): 27-43, 2015.

\bibitem[Ogurtsov et al.(2002)]{Ogurtsov}Ogurtsov, M.G., Nagovitsyn,
Y.A., Kocharov, G.E., Jungner, H.: Long-period cycles of the Sun\textquoteright s
activity recorded in direct solar data and proxies. Solar Physics
211, 371\textendash 394, 2002.

\bibitem[Owens and Forsyth(2013)]{Owen} Owens, M. J., Forsyth, R.
J.: The Heliospheric Magnetic Field. Living Rev. Solar Phys., 10,
5, 2013.

\bibitem[Pestiaux et al.(1988)]{Pestiaux}Pestiaux P., Duplessy, J.
C., van der Mersch, I., Berger, A.: Paleoclimatic variability at frequencies
ranging from 1 cycle per 10 000 years to 1 cycle per 1000 years: Evidence
for nonlinear behaviour of the climate system. Climatic Change, 12(1),
9\textendash 37, 1988.

\bibitem[Piovan and Milani(2006)]{Piovan} Piovan, L., Milani, F.:
Moto del Sole intorno al baricentro del Sistema Solare (Solar motion
around the solar system barycentre). Astronomia, 3, 38-44, 2006.

\bibitem[Phillips(2011)]{Phillips}Phillips, T.: Spacecraft sees solar
storm engulf Earth, Science@NASA, issued Aug 18, 2011.

\bibitem[Poluianovand Usoskin(2014)]{Poluianov2014}Poluianov, S.,
Usoskin, I.: Critical Analysis of a Hypothesis of the Planetary Tidal
Influence on Solar Activity. Solar Physics 289, 2333\textendash 234,
2014.

\bibitem[Potgieter(2013)]{Potgieter}Potgieter, M. S.: Solar Modulation
of Cosmic Rays. Living Rev. Solar Phys., 10, 3-66, 2013.

\bibitem[Puetz et al.(2014)]{Puetz}Puetz, S. J., Prokoph, A., Borchardt,
G., Mason, E. W.: Evidence of synchronous, decadal to billion year
cycles in geological, genetic, and astronomical events. Chaos Solitons
Fractals, 62\textendash 63, 55\textendash 75, 2014.

\bibitem[Reimer et al.(2004)]{Reimer}Reimer, P. J., Baillie, M. G.
L., Bard, E., Bayliss, A., Beck, J. W., Bertrand, C. J. H., Blackwell,
P. G., Buck, C. E., Burr, G. S., Cutler, K. B., Damon, P. E., Edwards,
R. L., Fairbanks, R. G., Friedrich, M., Guilderson, T. P., Hogg, A.
G., Hughen, K. A., Kromer, B., McCormac, G., Manning, S., Bronk Ramsey,
C., Reimer, R. W., Remmele, S., Southon, J. R., Stuiver, M., Talamo,
S., Taylor, F. W., Plicht, J. V. D., Weyhenmeyer, C. E.: IntCal04
terrestrial radiocarbon age calibration, 0-26 cal kyr BP. Radiocarbon,
46 (3), 1029\textendash 1058, 2004. https://www.radiocarbon.org/IntCal04.htm

\bibitem[Salvador(2013)]{Salvador}Salvador, R.: A Mathematical Model
of the Sunspot Cycle for the Past 1000 yr. Pattern Recognition in
Physics, 1, 117-122, 2013.

\bibitem[Scafetta et al.(2004)]{Scafetta2004}Scafetta, N., Grigolini,
P., Imholt, T., Roberts, J. A., West, B. J.: Solar turbulence in earth\textquoteright s
global and regional temperature anomalies. Physical Review E 69, 026303,
2004.

\bibitem[Scafetta and West(2006)]{ScafettaWest2006}Scafetta, N.,
West, B. J.: Phenomenological solar signature in 400 years of reconstructed
Northern Hemisphere temperature record. Geophysical Research Letters
33, L17718, 2006.

\bibitem[Scafetta(2009)]{Scafetta2009}Scafetta, N.: Empirical analysis
of the solar contribution to global mean air surface temperature change.
Journal of Atmospheric and Solar-Terrestrial Physics 71, 1916-1923,
2009.

\bibitem[Scafetta(2010)]{Scafetta2010}Scafetta, N.: Empirical evidence
for a celestial origin of the climate oscillations and its implications.
J. Atmos. Sol.-Terr. Phys., 72(13), 951\textendash 970, 2010.

\bibitem[Scafetta(2012a)]{Scafetta2012a}Scafetta, N.: Multi-scale
harmonic model for solar and climate cyclical variation throughout
the Holocene based on Jupiter\textendash Saturn tidal frequencies
plus the 11-year solar dynamo cycle. J. Atmos. Sol.- Terr. Phys.,
80, 296\textendash 311, 2012a.

\bibitem[Scafetta(2012b)]{Scafetta(2012b)}Scafetta, N.: Does the
Sun work as a nuclear fusion amplifier of planetary tidal forcing?
A proposal for a physical mechanism based on the mass-luminosity relation.
J. Atmos. Sol.-Terr. Phys., 81\textendash 82, 27\textendash 40, 2012b.

\bibitem[Scafetta(2012c)]{Scafetta2012c}Scafetta, N.: A shared frequency
set between the historical mid-latitude aurora records and the global
surface temperature. Journal of Atmospheric and Solar-Terrestrial
Physics 74, 145-163, 2012c.

\bibitem[Scafetta(2013)]{Scafetta2013}Scafetta, N.: Discussion on
climate oscillations: CMIP5 general circulation models versus a semi-empirical
harmonic model based on astronomical cycles. Earth-Sci. Rev., 126,
321\textendash 357, 2013.

\bibitem[Scafetta(2014a)]{Scafetta2014}Scafetta, N.: Discussion on
the spectral coherence between planetary, solar and climate oscillations:
a reply to some critiques. Astrophys. Space Sci., 354, 275-299, 2014a.

\bibitem[Scafetta(2014b)]{Scafettab}Scafetta, N.: The complex planetary
synchronization structure of the solar system. Pattern Recognition
in Physics 2, 1-19, 2014b. DOI: 10.5194/prp-2-1-2014.

\bibitem[Scafetta(2014c)]{Scafetta2014c}Scafetta, N.: Multi-scale
dynamical analysis (MSDA) of sea level records versus PDO, AMO, and
NAO indexes. Climate Dynamics, 43, 175-192, 2014c.

\bibitem[Scafetta(2016)]{Scafetta2016}Scafetta, N.: High resolution
coherence analysis between planetary and climate oscillations. Advances
in Space Research, 2016. DOI: 10.1016/j.asr.2016.02.029

\bibitem[Scafetta et al.(2013)]{ScafettaHumlum2013}Scafetta, N.,
Humlum, O., Solheim, J.-E, Stordahl, K.: Comment on \textquotedbl{}The
influence of planetary attractions on the solar tachocline\textquotedbl{}
by Callebaut, de Jager and Duhau. Journal of Atmospheric and Solar-Terrestrial
Physics 102, 368-371, 2013.

\bibitem[Scafetta and Willson(2013a)]{ScafettaWillson2013a}Scafetta,
N., Willson, R. C.: Planetary harmonics in the historical Hungarian
aurora record (1523\textendash 1960). Planet. Space Sci., 78, 38\textendash 44,
2013a. 

\bibitem[Scafetta and Willson(2013b)]{ScafettaandWillson(2013b)}Scafetta,
N., Willson, R. C.: Empirical evidences for a planetary modulation
of total solar irradiance and the TSI signature of the 1.09-year Earth\textendash Jupiter
conjunction cycle. Astrophys. Space Sci., 348(1), 25\textendash 39,
2013b.

\bibitem[Sharp(2013)]{Sharp}Sharp, G. J.: Are Uranus \& Neptune Responsible
for Solar Grand Minima and Solar Cycle Modulation? International Journal
of Astronomy and Astrophysics, 3, 260-273, 2013.

\bibitem[Shaviv et al.(2014)]{Shaviv}Shaviv, N. J., Prokoph, A.,
Veizer, J.: Is the Solar System\textquoteright s Galactic Motion Imprinted
in the Phanerozoic Climate? Scientific Reports 4, 6150, 2014. DOI:
10.1038/srep06150

\bibitem[Siegenthaler et al.(1980)]{Siegenthaler}Siegenthaler, U.,
Heimann, M., and Oeschger, H.: $^{14}C$ variations caused by changes
in the global carbon cycle. Radiocarbon, 22, 177\textendash 191, 1980.

\bibitem[Smythe and Eddy(1977)]{Smythe}Smythe, C.M., Eddy, J.A.,
1977. Planetary tides during maunder sunspot. Nature 266, 434\textendash 435,
1977.

\bibitem[Solanki et al.(2004)]{Solanki}Solanki, S.K., Usoskin, I.G.,
Kromer, B., Schussler, M., and Beer, J.: Unusual activity of the Sun
during recent decades compared to the previous 11,000 years, Nature,
431, 1084\textendash 1087, 2004.

\bibitem[Solheim(2013)]{Solheim}Solheim, J.-E.: Signals from the
planets, via the Sun to the Earth. Pattern Recogn. Phys., 1, 177-184,
2013.

\bibitem[Sonett(1984)]{Sonett1984}Sonett, C. P.: Very long solar
periods and radiocarbon record. Rev. Geophys. S. Phys., 22(3), 239\textendash 254,
1984.

\bibitem[Stefani et al.(2016)]{Stefani}Stefani, F., Giesecke, A.,Weber,
N., Weier, T.: Synchronized Helicity Oscillations: A Link Between
Planetary Tides and the Solar Cycle? Solar Physics (2016). doi:10.1007/s11207-016-0968-0

\bibitem[Stocker and Wright(1996)]{Stocker}Stocker, T. F., Wright,
D. G.: Rapid changes in ocean circulation and atmospheric radiocarbon.
Paleoceanography, 11, 773\textendash 795, 1996.

\bibitem[Steinhilber et al.(2009)]{Steinhilber}Steinhilber, F., Beer,
J., Fr{\"o}hlich, C.: Total solar irradiance during the Holocene.
Geophysics Research Letters 36, L19704, 2009.

\bibitem[Stuiver and Quay(1980)]{Stuiver1980}Stuiver, M., Quay, P.
D.: Changes in atmospheric carbon-14 attributed to a variable Sun.
Science, 207, 11\textendash 19, 1980.

\bibitem[Stuiver et al.(1998)]{Stuiver}Stuiver, M., Reimer, P. J.,
Bard, E., Beck, J. W., Burr, G. S., Hughen, K. A., Kromer, B., McCormac,
G., van der Plicht, J., Spurk, M.: IntCal98 radiocarbon age calibration,
24,000\textendash 0 cal BP. Radiocarbon 40(3), 1041\textendash 83,
1998.

\bibitem[Svensmark(1998)]{Svensmark}Svensmark, H.: Influence of Cosmic
Rays on the Earth's climate. Phys. Rev. Lett., 81, 5027-3030, 1998.

\bibitem[Svensmark et al.(2009)]{Svensmark2}Svensmark, H., Bondo,
T., Svensmark, J.: Cosmic ray decreases affect atmospheric aerosols
and clouds, Geophys. Res. Lett., 36, L15101, 2009. doi:10.1029/2009GL038429

\bibitem[Svensmark et al.(2012)]{Svensmark2012}Svensmark, J., Enghoff,
M. B., Svensmark, H.: Effects of cosmic ray decreases on cloud microphysics.
Atmos. Chem. Phys. Discuss., 12, 3595\textendash 3617, 2012. 

\bibitem[Tan and Cheng(2013)]{TanCheng}Tan, B., Cheng, Z.: The mid-term
and long-term solar quasiperiodic cycles and the possible relationship
with planetary motions. Astrophys. Space Sci., 343, 511\textendash 521,
2013.

\bibitem[Tattersall(2013a)]{Tattersall}Tattersall, R.: The hum: log-normal
distribution and planetary\textendash solar resonance. Pattern Recogn.
Phys., 1, 185\textendash 198, 2013a. 

\bibitem[Tattersall(2013b)]{Tattersalb}Tattersall, R.: Apparent relations
between planetary spin, orbit, and solar differential rotation. Pattern
Recognition Phys., 1, 199-202, 2013b.

\bibitem[Tinsley(2008)]{Tinsley}Tinsley, B. A.: The global atmospheric
electric circuit and its effects on cloud microphysics. Reports on
Progress in Physics, 71, 066801, 2008.

\bibitem[Usoskin et al.(2016)]{Usoskin}Usoskin, I. G., Gallet, Y.,
Lopes, F., Kovaltsov, G. A., Hulot, G.: Solar activity during the
Holocene: the Hallstatt cycle and its consequence for grand minima
and maxima. Astronomy \& Astrophysics 587, A150, 2016.

\bibitem[Vasiliev and Dergachev(1998)]{Vasiliev1998}Vasiliev, S.
S., Dergachev, V. A.: The change of natural radiocarbon level in the
Earth\textquoteright s atmosphere over the past 8000 years as a consequence
of solar activity, geomagnetic field and climatic factors: 2400-year
cycle. Biofizika, 43(4), 681\textendash 688, (in Russian), 1998.

\bibitem[Vasiliev and Dergachev(2002)]{Vasiliev}Vasiliev, S. S.,
Dergachev, V. A.: The 2400-year cycle in atmospheric radiocarbon concentration:
bispectrum of 14C data over the last 8000 years. Annales Geophysicae
20, 115\textendash 120, 2002.

\bibitem[Vaquero et al.(2002)]{Vaquero}Vaquero, J.M., Gallego, M.C.,
Garc{\'i}a, J. A.: A 250-year cycle in naked-eye observations of
sunspots. GRL, 29 (20), 58-1 58-4, 2002.

\bibitem[Wilson(2013)]{Wilson}Wilson, I. R. G.: The Venus\textendash Earth\textendash Jupiter
spin\textendash orbit coupling model. Pattern Recogn. Phys., 1, 147\textendash 158,
2013.

\bibitem[Wolf(1859)]{Wolf1859}Wolf, R.: Extract of a letter to Mr.
Carrington. Monthly Notices of the Royal Astronomical Society, 19,
85\textendash 86, 1859.

\bibitem[Wolff and Patrone(2010)]{Wolff}Wolff, C. L., Patrone, P.N.:
A new way that planets can affect the Sun. Sol. Phys., 266, 227\textendash 246,
2010.

\bibitem[Wyatt and Curry(2014)]{Wyatt}Wyatt, M. G., Curry, J. A.:
Role for Eurasian Arctic shelf sea ice in a secularly varying hemispheric
climate signal during the 20th century. Clim. Dynam., 42, 2763-2782,
2014.

\bibitem[Yu et al.(1983)]{Yu}Yu, Z., Chang, S., Kumazawa, M., Furumoto,
M., Yamamoto, A.: Presence of periodicity in meteorite falls. National
Institute of Polar Research 30, 362\textendash 366 Memoirs, Special
issue (ISSN 0386-0744), 1983.
\end{thebibliography}
\end{document}